\documentclass[12pt,twoside]{article}
\usepackage[latin1]{inputenc}
\usepackage{times}
\usepackage{epsfig}
\usepackage{amsfonts}
\usepackage{float}
\usepackage{amsmath}
\usepackage{latexsym}
\usepackage{graphicx}
\usepackage{nicefrac} 
\usepackage{type1cm}         
\usepackage{txfonts}
\usepackage{xcolor}

\relax

\newtheorem{definition}{{\bf Definition}\bf}[section]
\newcommand{\bdefi}{\begin{definition}}
\newcommand{\edefi}{\end{definition}}

\newtheorem{appropr}[definition]{{\bf Approximation Procedure}\bf}
\newcommand{\bappr}{\begin{appropr}}
\newcommand{\eappr}{\end{appropr}}

\newtheorem{bedi}[definition]{{\bf Condition}\bf}
\newcommand{\bbd}{\begin{bedi}}
\newcommand{\ebd}{\end{bedi}}

\newtheorem{bedin}[definition]{{\bf Conditions}\bf}
\newcommand{\bbdn}{\begin{bedin}}
\newcommand{\ebdn}{\end{bedin}}

\newtheorem{corollary}[definition]{{\bf Corollary}\bf}
\newcommand{\bco}{\begin{corollary}}
\newcommand{\eco}{\end{corollary}}

\newtheorem{lemma}[definition]{{\bf Lemma}\bf}
\newcommand{\blem}{\begin{lemma}}
\newcommand{\elem}{\end{lemma}}

\newtheorem{proposition}[definition]{{\bf Proposition}\bf}
\newcommand{\bpro}{\begin{proposition}}
\newcommand{\epro}{\end{proposition}}

\newtheorem{satz}[definition]{{\bf Theorem}\bf}
\newcommand{\bsa}{\begin{satz}}
\newcommand{\esa}{\end{satz}}

\newtheorem{assumption}[definition]{{\bf Assumption}\bf}
\newcommand{\bas}{\begin{assumption}}
\newcommand{\eas}{\end{assumption}}

\newtheorem{assumptions}[definition]{{\bf Assumptions}\bf}
\newcommand{\bass}{\begin{assumptions}}
\newcommand{\eass}{\end{assumptions}}

\newtheorem{abb}{{\bf Figure}\bf}
\newcommand{\babb}{\begin{abb}}
\newcommand{\eabb}{\end{abb}}

\newenvironment{remark}{\begin{rmk}\sl}{\end{rmk}}
\newtheorem{rmk}{{\bf Remark}\bf}[section]
\newcommand{\brem}{\begin{remark}}
\newcommand{\erem}{\end{remark}}

\newenvironment{remarks}{\begin{rmks}\sl}{\end{rmks}}
\newtheorem{rmks}{{\bf Remarks}\bf}[section]
\newcommand{\brems}{\begin{remarks}}
\newcommand{\erems}{\end{remarks}}

\newenvironment{example}{\begin{exmp}\rm}{\end{exmp}}
\newtheorem{exmp}{{\bf Example}\bf}[section]
\newcommand{\bbsp}{\begin{example}}
\newcommand{\ebsp}{\end{example}}
\newcommand{\bexa}{\begin{example}}
\newcommand{\eexa}{\end{example}}

\newtheorem{model}{{\bf Model}\bf}[section]
\newcommand{\bmdl}{\begin{model}}
\newcommand{\emdl}{\end{model}}

\newtheorem{scheme}{{\bf Scheme}\bf}[section]
\newcommand{\bscm}{\begin{scheme}}
\newcommand{\escm}{\end{scheme}}

\newenvironment{tabelle}{\begin{tabl}\sl}{\end{tabl}}
\newtheorem{tabl}{{\bf Table}\bf}
\newcommand{\btab}{\begin{tabelle}}
\newcommand{\etab}{\end{tabelle}}

\newenvironment{exercise}{\begin{exc}\sl}{\end{exc}}
\newtheorem{exc}{Exercise}[section]
\newcommand{\bexe}{\begin{exercise}}
\newcommand{\eexe}{\end{exercise}}

\newcommand{\df}{distribution function}
\newcommand{\db}{distribution}
\newcommand{\dfs}{distribution functions}
\newcommand{\dbs}{dis\-tri\-bu\-tions}

\newcommand{\rvs}{random variables}

\newcommand{\yp}{hypothesis}

\newcommand{\np}{non\-pa\-ra\-me\-tric}

\newcommand{\obs}{observation}
\newcommand{\obss}{observations}
\newcommand{\ci}{confidence interval}
\newcommand{\cis}{confidence intervals}

\newcommand{\asy}{asymptotic}
\newcommand{\ind}{independent}

\newcommand{\cm}{covariance matrix}
\newcommand{\cms}{covariance matrices}

\relax

\renewcommand{\theequation}{\mbox{\arabic{section}.\arabic{equation}}}

\newcommand{\dl}{\displaystyle}

\newcommand{\nnr}{\nonumber}
\newcommand{\ub}{\underbrace}

\newcommand{\sep}{\quad}

\newcommand{\mc}{\multicolumn}

\newcommand{\bay}{\begin{array}}
\newcommand{\eay}{\end{array}}

\newcommand{\bsl}{\begin{slide}}
\newcommand{\esl}{\end{slide}}

\newcommand{\bfr}{\begin{frame}}
\newcommand{\efr}{\end{frame}}

\newcommand{\bol}{\begin{overlay}}
\newcommand{\eol}{\end{overlay}}

\newcommand{\bqa}{\begin{eqnarray*}}
\newcommand{\eqa}{\end{eqnarray*}}

\newcommand{\bqan}{\begin{eqnarray}}
\newcommand{\eqan}{\end{eqnarray}}

\newcommand{\bqt}{\begin{quote}}
\newcommand{\eqt}{\end{quote}}

\newcommand{\bt}{\begin{tabbing}}
\newcommand{\et}{\end{tabbing}}

\newcommand{\bit}{\begin{itemize}}
\newcommand{\eit}{\end{itemize}}

\newcommand{\bist}{\begin{itemstep}}
\newcommand{\eist}{\end{itemstep}}

\newcommand{\ben}{\begin{enumerate}}
\newcommand{\een}{\end{enumerate}}

\newcommand{\beq}{\begin{equation}}
\newcommand{\eeq}{\end{equation}}

\newcommand{\bdes}{\begin{description}}
\newcommand{\edes}{\end{description}}

\newcommand{\btb}{\begin{tabular}}
\newcommand{\etb}{\end{tabular}}

\newcommand{\bpic}{\begin{picture}}
\newcommand{\epic}{\end{picture}}

\newcommand{\bcen}{\begin{center}}
\newcommand{\ecen}{\end{center}}

\newcommand{\bfg}{\begin{figure}}
\newcommand{\efg}{\end{figure}}

\newcommand{\bmp}{\begin{minipage}}
\newcommand{\emp}{\end{minipage}}

\newcommand{\bgan}{\begin{gather}}
\newcommand{\egan}{\end{gather}}

\newcommand{\bal}{\begin{align*}}
\newcommand{\eal}{\end{align*}}

\newcommand{\baln}{\begin{align}}
\newcommand{\ealn}{\end{align}}

\newcommand{\bala}{\begin{alignat*}}
\newcommand{\eala}{\end{alignat*}}

\newcommand{\balan}{\begin{alignat}}
\newcommand{\ealan}{\end{alignat}}

\newcommand{\bspt}{\begin{split}}
\newcommand{\espt}{\end{split}}

\newcommand{\bvt}{\begin{verbatim}}
\newcommand{\evt}{\end{verbatim}}

\newcommand{\lam}{\lambda}

\newcommand{\hF}{{\widehat F}}

\newcommand{\hH}{{\widehat H}}

\newcommand{\hlam}{{\widehat \lambda}}

\newcommand{\hsigma}{{\widehat \sigma}}

\newcommand{\htau}{{\widehat \tau}}

\newcommand{\htheta}{{\widehat \theta}}

\newcommand{\kA}{{\cal A}}
\newcommand{\kB}{{\cal B}}
\newcommand{\kC}{{\cal C}}
\newcommand{\kD}{{\cal D}}
\newcommand{\kE}{{\cal E}}
\newcommand{\kF}{{\cal F}}

\newcommand{\kS}{{\cal S}}

\newcommand{\kg}{k=1, \ldots,}

\newcommand{\sumi}{\sum_{i=1}^}
\newcommand{\sumj}{\sum_{j=1}^}
\newcommand{\sumk}{\sum_{k=1}^}
\newcommand{\sumell}{\sum_{\ell=1}^}
\newcommand{\sums}{\sum_{s=1}^}
\newcommand{\sumr}{\sum_{r=1}^}

\newcommand{\ks}{\oplus}

\newcommand{\var}{\operatorname{{\it Var}}}

\newcommand{\Cov}{\operatorname{{\it Cov}}}
\newcommand{\Var}{\operatorname{{\it Var}}}

\newcommand{\tr}{\operatorname{\it tr}}

\newcommand{\nfrac}{\nicefrac}

\relax

\newcommand{\olR}{\overline{R}}

\newcommand{\olY}{\overline{Y}}

\newcommand{\vA}{\boldsymbol{A}}
\newcommand{\vB}{\boldsymbol{B}}

\newcommand{\vI}{\boldsymbol{I}}
\newcommand{\vJ}{\boldsymbol{J}}

\newcommand{\vM}{\boldsymbol{M}}

\newcommand{\vP}{\boldsymbol{P}}

\newcommand{\vR}{\boldsymbol{R}}

\newcommand{\vV}{\boldsymbol{V}}

\newcommand{\veins}{{\bf 1}}
\newcommand{\vnull}{{\bf 0}}

\newcommand{\wtN}{\widetilde{N}}

\newcommand{\wtsigma}{\widetilde{\sigma}}

\begin{document}

\thispagestyle{empty}
\begin{center}
{\Large \textbf{An unbiased rank-based estimator of the Mann-Whitney variance including the case of ties}}
\end{center}
\bcen
{\large Edgar Brunner} \\[1ex]
{\it Department of Medical Statistics, University of G{\"o}ttingen, Germany} 
\\[2ex]
{\large Frank Konietschke}\\[1ex]
{\it Institute of Biometry and Clinical Epidemiology, Charite Berlin, Germany 
} \\[2ex]

\ecen
\mbox{ }\\
\bcen
{\sc Abstract} \\[3mm]
\ecen
Many estimators of the variance of the well-known unbiased and uniform most powerful estimator $\htheta$ of the Mann-Whitney effect, $\theta = P(X < Y) + \nfrac12 P(X=Y)$, are considered in the literature. Some of these estimators are only valid in case of no ties or are biased in case of small sample sizes where the amount of the bias is not discussed. Here we derive an unbiased estimator that is based on different rankings, the so-called 'placements' (Orban and Wolfe, 1980), and is therefore easy to compute. This estimator does not require the assumption of continuous \dfs\ and is also valid in the case of ties. Moreover, it is shown that this estimator is non-negative and has a sharp upper bound which may be considered an empirical version of the well-known Birnbaum-Klose inequality. The derivation of this estimator provides an option to compute the biases of some commonly used estimators in the literature. Simulations demonstrate that, for small sample sizes, the biases of these estimators depend on the underlying \dfs\ and thus are not under control. This means that in the case of a biased estimator, simulation results for the type-I error of a test or the coverage probability of a \ci\ do not only depend on the quality of the approximation of $\htheta$ by a normal \db\ but also an additional unknown bias caused by the variance estimator. Finally, it is shown that this estimator is $L_2$-consistent. \\[4ex]
\newpage
\mbox{ }

\renewcommand{\baselinestretch}{1}




\section{Introduction}\label{sec1}
The Mann-Whitney effect $\theta = P(X_1 \le X_2)$ is a common nonparametric effect that is used to describe a treatment effect in a nonparametric setting for two independent random variables $X_1 \sim F_1(x)$ and $X_2 \sim F_2(x)$ involving continuous \dfs. 
This effect was introduced by Mann and Whitney (1947) for testing the \yp\ $H_0^F: F_1 = F_2$. A few years later, Putter (1955) considered the case of ties in some \np\ tests and derived the consistency region for the Mann-Whitney test in the general case as $\theta = P(X_1 < X_2) + \frac12 P(X_1=X_2) \neq \frac12$. 
An unbiased and $L_2$-consistent estimator $\htheta$ of the Mann-Whitney effect $\theta$ can be obtained from $U$-statistics theory and it is well known (see, e.g., Lehmann (1951) that this estimator is the uniform most powerful unbiased estimator of $\theta$. Moreover, it can be represented by ranks. To compute \cis\ for $\theta$, the variance $\sigma_N^2$ of this estimator $\htheta$ is required and it seems to be less simple to provide an unbiased estimator of $\sigma_N^2$ which is also valid in case of ties. 

Under the assumption of continuous \dfs\ Sen (1967) and Govindarajulu (1968) provided unbiased estimators of $\sigma_N^2$ but did not discuss whether these estimators could become negative. Hilgers (1981) derived a rank representation of Sen's estimator but did not discuss whether this estimator could become negative. Shirahata (1993) derived an unbiased estimator of the Mann-Whitney variance based on U-statistics which is only valid in case of no ties. Shirahata mentioned that his estimator is equivalent to Sen's estimator but he also mentions that there is a possibility that it might become negative. In his simulations, however, he did not observe negative values. In Sect.~\ref{Biaslit} it will be shown that this estimator is non-negative in case of no ties - disproving Shirahata's conjecture. But, when in the case of ties, the ranks in Hilgers' estimator are replaced with mid-ranks, it turns out by simple counterexamples that this estimator can become negative. This shows that it is not trivial to derive an unbiased and non-negative estimator of the Mann-Whitney variance.

It seems that Bamber (1975) was the first to provide (without detailed proof) an unbiased estimator which is also valid in the case of ties. On the one hand, the representation of this estimator is quite involved. It is a linear combination of several positive and negative terms and thus, it is unclear whether this estimator might become negative. This is not discussed in Bamber's paper. Moreover, this estimator is not well recognized in the statistical literature and recent publications (see, e.g., Brunner and Munzel (2000), Perme and Manevski (2019), Gasparyan et al. (2021) mainly use the so-called DeLong-estimator, $\hsigma_{DL}^2$ (DeLong et al. 1988) which can conveniently be represented using ranks and is also valid in case of ties. DeLong's estimator is non-negative since it can be written as a weighted sum of quadratic forms involving only non-negative weights and using certain rankings. For small sample sizes, however, $\hsigma_{DL}^2$ can be biased. Some details on variance estimators from the literature are discussed in Sect.~\ref{Biaslit}.  

Therefore, it shall be one of the aims of this paper to provide a simple representation of an unbiased estimator of $\sigma_N^2$ which is also valid in case of ties. Moreover, some basic properties of this estimator such as non-negativity, $L_2$-consistency, and an empirical version of the Birnbaum-Klose inequality (1957) will be investigated.

The paper is organized as follows. The statistical model and some basic notations are stated in Sect.~\ref{statmod} where the exact (finite) variance of the Mann-Whitney statistic involving the case of ties is briefly considered {\color{black} to provide the necessary notations}. An unbiased estimator of the Mann-Whitney variance is derived in Sect.~\ref{deriv} by {\color{black} deriving} the covariance matrix of the placements in both samples {\color{black} which provides the key to compute the bias of an estimator using Lancaster's theorem}. In Sect.~\ref{Biaslit} some well-known estimators from the literature and their properties are investigated. The proofs of the main results are deferred to the Appendix.

\section{Statistical model and notations} \label{statmod}

\subsection{Mann-Whitney estimator and placements} \label{MWplace}
Let $X_{ik} \sim F_i$, $\kg n_i$; $i=1,2$, be \ind\ and identically distributed \rvs. Let further $\theta = P(X_{1k} < X_{2r}) + \frac12 P(X_{1k} = X_{2r}) \ = \ \int F_1 d F_2$ denote the Mann-Whitney effect 
which can equivalently be written as
\bqan
 \theta &=& \int F_1 d F_2 \hspace*{10ex} \text{and \sep} 1 - \theta \ = \ \int 
            F_2 d F_1 \ ,  \label{thetadef}
\eqan
where $F_i(x) = \frac12 \left[ F_i^+(x) + F_i^-(x) \right]$ denotes the so-called normalized version of the \df\ {\color{black} in the sense of Lévy (1925)}. Here, $F_i^-(x)$ denotes the left-continuous and $F_i^+(x)$ the right-continuous version of the \df\ (Ruymgaart, 1980). {\color{black} For further explanations regarding this version of the \df\ we refer to Brunner and Puri (2001, Sects.1.2.2 and 1.3.1), Brunner and Puri (2002, Sect.2 and Appendix, Lemma A.1), and Kruskal (1952, Sect.9).}

A simple plug-in estimator of $\theta$ is given by
\bqan
 \htheta = \int \hF_1 d\hF_2 &=& \frac1{n_2} \sumk {n_2} \hF_1 (X_{2k}) \nnr \\[1ex]
  &=& \frac1{n_1} \left(\olR_{2 \cdot} - \frac{n_2+1}2 \right) \ = \ \frac1N \left(\olR_{2 \cdot} - \olR_{1 \cdot} \right)+\frac12 \ , \label{f12plugin}
\eqan
{\color{black} where $\hF_i(x)$ denotes the normalized version of the empirical \df\ (see Def.2.1.3 in Brunner, Bathke, Konietschke, 2019)} and $\olR_{i \cdot} = \frac1{n_i} {\color{black} \sumk {n_i}} R_{ik}$, $i=1,2$, {\color{black} denotes the mean of the ranks $R_{ik}$ of $X_{ik}$ in sample $i$ among all $N = n_1 + n_2$ \obss\ (overall rank). Note that the last step in (\ref{f12plugin}) follows from $R_{1\cdot} + R_{2\cdot} = N(N+1)/2$ (see also Result~3.1 and Exercise/Problem~3.7 in Brunner et al., 2019).} 
It is well-known that $\htheta$ is an unbiased and $L_2$-consistent estimator of $\int F_1 d F_2$, {\color{black} which can immediately be seen from Theorem~\ref{varhtheta}}. \\[1ex] 
{\color{black} Let 
\bqan 
c(x,y) &=& \left\{\bay{rl} 0 &, x > y \\ \nfrac12 &, x=y \\ 1 &, x<y \eay \right.  \label{count}
\eqan 
denote the indicator function - also denoted as 'counting statistic' (Randles and Wolfe, 1979).  Then, }
\bqa
 E(\htheta) &=& E \left( \int \hF_1 d\hF_2 \right) \ = \ E \left( \frac1{n_2} 
                \sumk {n_2} \hF_1 (X_{2k}) \right) \\
	&=& \frac1{n_1n_2} \sumr {n_1} \sumk {n_2} E [c( X_{1r}, X_{2k}) ] 	\ = \ \int F_1 dF_2 \ ,
\eqa
(for details we refer, e.g., to Brunner et al. 2019, Sect.~7.2.3). We note that in most textbooks, $\htheta$ is defined as 
\bqan
 \htheta &=& \frac1{n_1n_2} \sumr {n_1} \sumk {n_2} c(X_{1r}, X_{2k}) \ , \label{f12count}
\eqan
which is identical to (\ref{f12plugin}). We prefer the approach by using the empirical \dfs, for convenience. The results are identical to those obtained using the indicator function $c(x,y)$. The relation of $\htheta$ in (\ref{f12count}) to (\ref{f12plugin}) follows immediately from the definition of the empirical \df\ $\hF_1(x) = 
\frac1{n_1} \sumr {n_1} c(X_{1r}, x)$ and the relation $n_1 \hF_1 (X_{2k}) = R_{2k} - R_{2k}^{(2)}$, where $R_{2k}^{(2)}$ is the rank of $X_{2k}$ among all $n_2$ \obss\ within sample~2. In general, the {\it overall rank} $R_{ik}$ of an \obs\ $X_{ik}$, \ $\kg n_i$, \ $i=1,2$, is defined as $R_{ik} = \frac12 + \sumj 2 \sumr {n_j} c(X_{jr}, X_{ik})$ and the {\it internal rank} of $X_{ik}$ within sample $i$ is defined as $R_{ik}^{(i)} = \frac12 + \sumr {n_i} c(X_{ir}, X_{ik})$. Note that $\nfrac12$ must be added to the sum of the indicator functions to obtain the intuitive meaning of ranks as place numbers of the \obss\ in the order statistic in case of no ties since the comparison $c(X_{ik}, X_{ik})$ of the \obs\ $X_{ik}$ with itself equals $\nfrac12$.

The quantities 
\bqan 
R_{1k}^* &=& R_{1k} - R_{1k}^{(1)} = n_2 \hF_2(X_{1k}) \ =\ \sumell {n_2} {\color{black} c(X_{2 \ell}, X_{1k}) } \ , \label{r1kst} \\
R_{2\ell}^* &=& R_{2\ell} - R_{2\ell}^{(2)} = n_1 \hF_1(X_{2\ell}) \ = \ \ \sumk {n_1} {\color{black} 
c(X_{1k}, X_{2 \ell}) } \label{r2lst}
\eqan 
are called 'placements' (Orban and Wolfe, 1980, 1982). The placements $R_{1k}^*$ and $R_{2\ell}^*$ are basically the place numbers (ranks) of $X_{1k}$ within sample~2 and of $X_{2\ell}$ within sample~1. The formal representations in (\ref{r1kst}) and (\ref{r2lst}) are immediately obvious from the following considerations. Let $\hH(x) = \frac1N [n_1 \hF_1(x) + n_2 \hF_2(x)]$ denote the weighted mean of the empirical \dfs\ $\hF_1(x)$ and $\hF_2(x)$. Then
\bqa 
\ub{N \hH(X_{1k})}_{R_{1k}-\frac12} &=& \sumj 2 \sums {n_j} c(X_{js}, X_{1k}) \ = \ n_1 \hF_1(X_{1k}) + n_2 \hF_2(X_{1k}) \\
&=& \ub{ \sums {n_1} c(X_{1s}, X_{1k}) }_{R_{1k}^{(1)}-\frac12} \ + \ \ub{\sums {n_2} c(X_{2s}, X_{1k}) }_{R_{1k}^*} 
\eqa 
and the expression for $R_{2\ell}^* $ follows in the same way. Then the mean  ${\color{black}\frac1{n_1n_2}} \sumell {n_2} R_{2\ell}^* = \frac1{n_1} \olR_{2 \cdot}^*= \frac1{n_1} (\olR_{2 \cdot} - \frac{n_2+1}2)$ yields the representation of $\htheta$ in (\ref{f12plugin}).

\subsection{The variance of the Mann-Whitney estimator} \label{MWvar}

First, we briefly derive the variance $\sigma_N^2$ of $\htheta$ including the case of discrete \dbs\ since some quantities appearing in this derivation will be used later to derive an unbiased estimator of $\sigma_N^2$. We note that the representation of the Mann-Whitney variance $\sigma_N^2$ presented below is identical to that in Bamber (1975) which extends the representation for continuous \dfs\ by van Dantzig (1951) to the case of ties. 

Since $E(\htheta) = \theta$, the variance is given by $\Var(\htheta) = E\left( \htheta - \theta \right)^2$ and it follows that
\bqan
 (\htheta - \theta)^2 &=& \left( \frac1{n_2} \sumk {n_2} \hF_1(X_{2k}) - \theta  \right)^2 \ = \ \left( \frac1{n_2} \sumk {n_2} \left[ \hF_1(X_{2k}) - \theta  \right] \right)^2 \nnr\  \\[1ex]
  &=& \frac1{n_1^2n_2^2} \sumk {n_2} \sumell {n_2} \sumr {n_1} \sums {n_1} \left[c(X_{1r}, X_{2k})-\theta\right]  
      \left[c(X_{1s}, X_{2\ell})-\theta \right] \ . \label{sdqathdq}
\eqan

When taking the expectation of $(\htheta - \theta)^2$, four cases in (\ref{sdqathdq}) must be distinguished. They are given in the following table where the expectations of the products of the indicator function in (\ref{sdqathdq}) are listed in the right column and are obtained by computing conditional expectations using routine computations. The numbers of combination cases (left column) are listed in the middle column. \\[2ex]
\btb{llcccl}
\mc{2}{c}{Combination} &  \mc{2}{c}{Number of Cases} &  & $E\left(\left[c(X_{1r}, X_{2k})-\theta\right]  \left[c(X_{1s}, X_{2\ell})-\theta \right] \right)$ \\ 
$k=\ell$,      & $r=s$  &   & $n_1n_2$  & & $ = \ \theta(1-\theta) - \frac14 \int (F_1^+ -F_1^-)dF_2$ \\[1.5ex]
$k=\ell$,      & $r\neq s$ & & $n_1n_2(n_1-1)$ & & $ = \ \int F_1^2 dF_2 - \theta^2 \ = \ \sigma_2^2$ \\[1.5ex]
$k \neq \ell$, & $r=s$     & & $(n_2-1)n_1n_2$ & & $ = \ \int F_2^2 dF_1 - (1-\theta)^2 \ = \ \sigma_1^2$ \\[1.5ex]
$k \neq \ell$, & $r\neq s$ & & $(n_1-1)n_1(n_2-1)n_2$ & & $ = \ 0$ \\ 
\etb
\text{ } \\[2ex]
The result of the summation over all combinations is given in the following theorem. \\[1ex]

\bsa \label{varhtheta}
The variance $\sigma_N^2$ of $\htheta$ is given by \\[-2ex]
\bqan
 \sigma_N^2 &=& \frac1{n_1n_2} \left[ (n_2-1) \sigma_1^2 + (n_1-1) \sigma_2^2 
							  + \theta(1-\theta) - \frac14 \tau \right], \label{repsignq} \\[-3ex]
         \nnr\ 
\eqan
where \\[-4ex]
\bqan
\sigma_1^2 &=& \Var\left(F_2(X_{11})\right) = \int F_2^2dF_1 - (1-\theta)^2 \ , \label{sig1q}  \\ 
\sigma_2^2 &=& \Var\left(F_1(X_{21})\right) = \int F_1^2dF_2- \theta^2 \ , \label{sig2q}  \\ 
\tau &=& \int (F_1^+ - F_1^-)dF_2 \ = \ \int (F_2^+ - F_2^-)dF_1 \ = \ P(X_{11} = X_{21}) \ . \label{betadef}
\eqan
\esa

\begin{remark}
 The quantity $\tau$ in (\ref{betadef}) can easily be interpreted as the probability of ties in the overlap region of $F_1$ and $F_2$.
\end{remark}

\section{Derivation of an unbiased estimator of $\sigma_N^2$} \label{deriv}

In this section, we derive an estimator of $\sigma_N^2$ which is unbiased and non-negative for all sample sizes $n_1, n_2 \ge 2$ and is also valid in case of ties. Moreover, we give some important properties of this estimator in Theorem~\ref{varhthetaest}. First, some notations and relations between sums of count functions and different rankings are considered. 

\subsection{Notations and Basic Results} \label{basicres} 
First, we note that 
\bqan
 c^2(x,y) &=& c(x,y) - \tfrac14 [ c^+(x,y) - c^-(x,y)] , \label{cqdef}
\eqan 
where $c^+(\cdot)$ denotes the right-continuous version and $c^-(\cdot)$ the left -continuous version of the indicator function. In particular, 
\bqan 
 E\left[c^2(X_{2\ell}, X_{1k}) \right] &=& \int F_2 dF_1 - \tfrac14 \tau \ = \ 1 - \theta - \tfrac14 \tau , \label{ecqxy}
\eqan 
where $\tau$ is defined in (\ref{betadef}).

Since $\htheta$ in (\ref{thetadef}) is the mean of the placements $R_{2\ell}^*$ as considered in Sect.~\ref{MWplace}, the placements $R_{1k}^*$ and  $R_{2\ell}^*$ shall be considered in more detail. They are the empirical counterparts of the quantities {\color{black} $n_2 F_2(X_{1k})$ and $n_1 F_1(X_{2\ell})$} and  will be used to estimate the variances $\sigma_1^2$ and $\sigma_2^2$ in (\ref{sig1q}) and (\ref{sig2q}). 

To compute the variance of $\htheta = \frac1{n_2} \sumell {n_2} \hF_1(X_{2\ell})$, the \cms\ of the {\color{black} scaled placement vectors $\frac1{N-n_i} \vR_i^* =\frac1{N-n_i} (R_{i1}^*, \ldots, R_{in_i}^*)'$}, $i=1,2$, shall be derived. Since $E[ c(X_{2\ell}, X_{1r}) ] =  \int F_2 dF_1 = 1 - \theta$ and $E[c(X_{1k}, X_{2r})] = \int F_1 dF_2 = \theta$, it follows from (\ref{r1kst}) and (\ref{r2lst}) that the expectations of the placement vectors are $E(\vR_1^*) = n_2 (1 - \theta) \veins_{n_1}$ and $E(\vR_2^*) = n_1 \theta \veins_{n_2}$. 

The structure of the \cms\ $\vV_i = \Cov(\vR_i^*)$, $i=1,2$ follows from the relations in (\ref{r1kst}) and (\ref{r2lst}) by noting that the $X_{ik}$ are \ind\ and identically distributed within each group $i$, by assumption. Thus, the common \df\ of the $X_{i1}, \ldots, X_{in_i}$ is invariant under all permutations and the variances of $R_{ik}^* = n_r \hF_r(X_{ik})$, \ $i \neq r = 1,2$ are all the same, $s_i^2$, say. Moreover, all covariances $\rho_i = \Cov(R_{ik}^*, R_{i\ell}^*)$ are identical for $k \neq \ell = 1, \ldots, n_i, \ i=1,2$, and it follows that $\vV_i$ has a compound symmetry structure,
\bqan
\vV_i \ = \ \Cov(\vR_i^*) &=& (s_i^2 - \rho_i) \vI_{n_i} + \rho_i \vJ_{n_i}, \ i=1,2,  \label{csvi}
\eqan
{\color{black} where $\vI_{n_i}$ denotes the unit matrix of dimension $n_i$ and $\vJ_{n_i} = \veins_{n_i} \veins_{n_i}'$ denotes the $(n_i \times n_i)$-matrix of 1s.}

By the same arguments, the covariances $\Cov(R_{1k}^*, R_{2\ell}^*)$ are all identical, $\rho_3$, say, and the total \cm\ $\vV$ is given by
\bqan
 \vV \ = \ \Cov \left(\bay{c} \vR_1^* \\ \cdots \\ \vR_2^* \eay \right) &=&
         \left(\bay{ccccc} (s_1^2 - \rho_1) \vI_{n_1} + \rho_1 \vJ_{n_1} & & \vdots & & \rho_3 \vJ_{n_1 \times n_2} \\ \mc{5}{c}{\cdots\cdots\cdots\cdots\cdots\cdots\cdots\cdots\cdots\cdots\cdots\cdots\cdots\cdots\cdots\cdots} \\[-2ex]
					\rho_3 \vJ_{n_2 \times n_1} & & \vdots & & (s_2^2 - \rho_2) \vI_{n_2} + \rho_2 \vJ_{n_2} \eay \right) 
                    \nnr\ \\[1ex] 
					&=& \left(\bay{ccccc} \vV_1 & & \vdots & & \rho_3 \vJ_{n_1 \times n_2} \\ 
                        \mc{5}{c}{\cdots\cdots\cdots\cdots\cdots\cdots\cdots\cdots} \\[-2ex]
					\rho_3 \vJ_{n_2 \times n_1} & & \vdots & & \vV_2 \eay \right) \label{3.v} \ .
\eqan

To derive an unbiased estimator of $\sigma_N^2$ in (\ref{repsignq}), we consider the quadratic forms 
\bqan
Q_i^2 &=& \sumk {n_i} \left( R_{ik}^* - \olR_{i \cdot}^* \right)^2, \ i=1,2  \label{qi2qf}
\eqan
of the centered placements $R_{ik}^* - \olR_{i \cdot}^* = R_{ik} - R_{ik}^{(i)} - \left( \olR_{i \cdot} - \tfrac{n_i+1}2 \right)$. 

\brems\ \text{ }
\ben 
 \item Many estimators of $\sigma_N^2$ from the literature can be written as a function of $Q_1^2$ and $Q_2^2$ {\color{black} (for details see Sect.~\ref{Biaslit}, Table~\ref{estimators})}. 
 \item In this case, the representation of the \cm\ of $\dl \vR^* = \left( \bay{c} \vR_1^* \\ \cdots \\ \vR_2^* \eay \right)$ enables a simple computation of the expectation of an estimator of $\sigma_N^2$. 
\een 
\erems 

Let $\vP_{n_i} = \vI_{n_i} - \frac1{n_i} \vJ_{n_i}$ denote the $n_i$-dimensional centering matrix.  Then, $Q_i^2$ can be written as $Q_i^2 = (\vR_i^*)' \vP_{n_i} \vR_i^*$ and by Lancaster's theorem, 
\bqan
 E(Q_i^2) &=& \tr (\vP_{n_i} \vV_i) + E \left[ (\vR_i^*)' \right] \vP_{n_i} E \left[\vR_i^* \right] \nnr\ \\
 &=& \tr (\vP_{n_i} \vV_i)  \label{qu2lanc}
\eqan
since $n_2^2 (1 - \theta)^2 \veins_{n_1}' \vP_{n_1} \veins_{n_1} = 0$ and $n_1^2 \theta^2 \veins_{n_2}' \vP_{n_2} \veins_{n_2} = 0$.

{\color{black} Note that from (\ref{csvi}), $\vP_{n_i} \vV_i = [s_i^2 - \rho_i] \vP_{n_i}$, since $\vP_{n_i} \vJ_{n_i} = \vnull$. Then, from $\tr (\vP_{n_i}) = n_i-1$, it follows from (\ref{csvi}) that
}
\bqan 
 \tr (\vP_{n_i} \vV_i) &=&  (n_i-1) [s_i^2 - \rho_i]; \ i=1,2 . \label{EQI}
\eqan 
Only the coefficients $s_1^2$, $s_2^2$ and $\rho_1, \rho_2$, and $\rho_3$ have to be determined. They are obtained from
\bqan
 s_1^2 &=& \Var\left( R_{1k} - R_{1k}^{(1)} \right) \ = \ \Var \left( 
             \sumell {n_2}  c(X_{2\ell}, X_{1k}) \right) \nnr\ \\ 
 			&=& E \left[\sumell {n_2} \sumr {n_2} [c(X_{2\ell}, X_{1k})-(1-\theta)] 
			    [c(X_{2r}, X_{1k}) -(1-\theta)] \right]  \label{varpl1} \\
 s_2^2 &=& \Var\left(R_{2k} - R_{2k}^{(2)}\right) \ = \ \Var \left( 
              \sumell {n_1}  c(X_{1\ell}, X_{2k}) \right) \nnr\ \\ 
			&=& E \left[\sumell {n_1} \sumr {n_1} [c(X_{1\ell}, X_{2k}) -\theta] 
			    [ c(X_{1r}, X_{2k}) - \theta] \right]   \label{varpl2} \\
 \rho_1 &=& \Cov \left(R_{11}-R_{11}^{(1)}, R_{12}-R_{12}^{(1)} \right) \nnr\ \\
         &=& E \left[\sumell {n_2} \sumr {n_2} [c(X_{2\ell}, X_{11}) - \theta] 
				     [c(X_{2r}, X_{12})-\theta]  \right] \label{covlks} \\
 \rho_2 &=& \Cov \left(R_{21}-R_{21}^{(2)}, R_{22}-R_{22}^{(2)} \right) \nnr\ \\
     &=& E \left[\sumell {n_1} \sumr {n_1} [c(X_{1\ell}, X_{21}) - (1-\theta)] 
		     [c(X_{1r}, X_{22}) -(1-\theta)] \right] \ .  \label{cov2ellr} 
\eqan

The expectations are obtained from Section~\ref{MWvar} by some routine computations and using (\ref{cqdef}). This leads to
\bqan
 s_1^2 \ = \ n_2 \left[(n_2-1) \sigma_1^2 + \theta(1-\theta) - \frac14 \tau 
 \right] & \text{and} & \rho_1 \ = \ n_2 \sigma_2^2  \label{tau1qc1} \\
 s_2^2 \ = \ n_1 \left[(n_1-1) \sigma_2^2 + \theta(1-\theta) - \frac14 \tau 
 \right]  & \text{and} & \rho_2 \ = \ n_1 \sigma_1^2 \label{tau2qc2} \ ,
\eqan
where $\tau$ is defined in (\ref{betadef}). Finally, the expectations of $Q_1^2$ and $Q_2^2$ are 
\bqan 
 E(Q_1^2) &=& (n_1-1) n_2 \left[(n_2-1) \sigma_1^2 - \sigma_2^2 + \theta(1-\theta) - \frac14 \tau \right] ,    
               \label{EQ1qdetail} \\
 E(Q_2^2) &=& (n_2-1) n_1 \left[(n_1-1) \sigma_2^2 - \sigma_1^2 + \theta(1-\theta) - \frac14 \tau \right] . 
               \label{EQ2qdetail}
\eqan 

Since the expectations of $Q_1^2$ and $Q_2^2$ are mixtures of all the quantities $s_1^2, s_2^2, \rho_1$, and $\rho_2$, it is preferable to consider the expectation of the sum $Q_1^2 + Q_2^2$. 

Now let $K(\theta, \tau) = \theta(1-\theta) - \frac14 \tau$, for convenience {\color{black} and let $\ks$ denote the direct sum of matrices and let $\odot$ denote the Hadamard product of matrices}. Then it follows from~(\ref{csvi}) and (\ref{3.v}) {\color{black} and from $\vV = \vV'$} that 
\bqa
E(Q_1^2 + Q_2^2) &=& \tr \left[\left(\vP_{n_1} \ks \vP_{n_2}\right) \vV \right] \ = \ \veins_N' \left[ \left( 
                     \vP_{n_1} \ks \vP_{n_2} \right) {\color{black} \odot} \vV \right] \veins_N \\
	&=& \sumi 2 \veins_{n_i}' (\vP_{n_i} {\color{black} \odot} \vV_i) \veins_{n_i} \ = \ \sumi 2 (n_i-1) [s_i^2 - \rho_i] \\
    &=& (n_1-1)n_2 \left[(n_2-1)\sigma_1^2-\sigma_2^2 + K(\theta, \tau) \right] \\
    & & + (n_2-1)n_1 \left[(n_1-1)\sigma_2^2-\sigma_1^2 + K(\theta, \tau) \right] \\
    &=& (n_2-1)[n_1n_2-N] \sigma_1^2 + (n_1-1)[n_1n_2-N] \sigma_2^2 \\	
    & &  + [n_1n_2-N + n_1n_2] K(\theta, \tau) \\
    &=& (n_1n_2-N) \ub{\left[ (n_2-1) \sigma_1^2 + (n_1-1) \sigma_2^2 + K(\theta,\tau) \right]}_{n_1n_2 \sigma_N^2} 
         + n_1n_2 K(\theta, \tau) \ ,
\eqa
which simplifies to
\bqan
  E \left( \frac1{n_1n_2} (Q_1^2+Q_2^2) - [ \theta(1-\theta) - \tfrac14 \tau ] \right) &=& (n_1n_2-N) \sigma_N^2 \ .
\label{eq1q2}
\eqan

Now replacing $\theta(1-\theta)$ by $E[ \htheta (1-\htheta) + \sigma_N^2] = \theta(1-\theta)$ finally leads to 
\bqan
 E \left( \frac1{n_1n_2} (Q_1^2+Q_2^2) - \left[ \htheta(1-\htheta) - \tfrac14 \tau \right] \right) 
   &=& (n_1n_2-N+1) \sigma_N^2 \nnr\ \\
   &=& (n_1-1)(n_2-1) \sigma_N^2 . \label{hsigmaNtau}
\eqan

It remains to find an unbiased estimator of $\tau = P(X_{11} = X_{21})$. To this end let $F_1^+(x) = P(X_{11} \le  x)$ and $F_1^-(x) = P(X_{11} <  x)$. Then it holds that $E[F_1^+(X_{21})] = P(X_{11} \le  X_{21})$ and $E[F_1^- (X_{21})] = P(X_{11} <  X_{21})$. Thus, 
\bqa
E\left[ F_1^+(X_{21}) - F_1^-(X_{21})  \right] &=& P(X_{11} \le  X_{21}) - P(X_{11} <  X_{21}) \\
&=& P(X_{11} = X_{21}) \ .
\eqa
To estimate these quantities let $c^+(y,x) {\color{black} =} \left\{ \bay{l} 1, \ y \le x, \\ 0, \ y>x \eay \right.$ and $c^-(y,x) {\color{black} =} \left\{ \bay{l} 1, \ y < x, \\ 0, \ y \ge x \eay \right.$ denote the right- and left-continuous versions of the indicator function, respectively. Further let $\hF_1^+(x) = \frac1{n_1} \sumr {n_1} c^+(X_{1r},x)$ denote the right-continuous version and $\hF_1^-(x) = \frac1{n_1} \sumr {n_1} c^-(X_{1r},x)$ the left-continuous version, respectively, of the empirical \df\ of $F_1$. Using Relation (\ref{r2lst}) it follows that the plug-in estimator 
\bqan
 \htau_N &=& \frac1{n_2} \sumell {n_2} \left[\hF_1^+(X_{2 \ell}) - \hF_1^-(X_{2 \ell}) \right] \nnr\ \\
         &=& \frac1{n_2} \sumell {n_2} \frac1{n_1} \left[\left( R_{2\ell}^+ - R_{2\ell}^- \right) - \left( \olR_{2 \ell}^{(2) +} - \olR_{2 \ell}^{(2) -} \right) \right] \nnr\ \\
         &=& \frac1{n_1} \left[\olR_{2 \cdot}^{\ +} - \olR_{2 \cdot}^{\ -} - \left( \olR_{2 \cdot}^{\ (2) +} - \olR_{2 \cdot}^{\ (2) -} \right)  \right].  \label{tauNest}
\eqan 
is an unbiased estimator of $\tau$. Here, $R_{2\ell}^+$ and $R_{2\ell}^-$ denote the maximal and minimal overall ranks of $X_{2\ell}$, respectively, and $R_{2\ell}^{(2) +}$ and $R_{2\ell}^{(2) -}$ denote the maximal and minimal internal ranks of $X_{2\ell}$ within sample~2. For details regarding maximal and minimal ranks, we refer to Brunner et al. (2019), Section~2.3.2, Def.~2.19. Unbiasedness of $\htau_N$ follows by noting that $E\left[ \hF_1^+(X_{21}) \right] = P(X_{11} \le  X_{21})$ and $E\left[ \hF_1^-(X_{21}) \right] = P(X_{11} < X_{21})$ and in turn $E\left[ \hF_1^+(X_{21}) - \hF_1^-(X_{21}) \right] = \tau$. 

{\color{black} Replacing $\tau$ in (\ref{hsigmaNtau}) with the unbiased estimator $\htau$ leads to the final result stated in Theorem~\ref{varhthetaest} where the foregoing derivations are summarized and some other important properties of the estimator $\hsigma_N^2$ in (\ref{sNquadrat}) are stated.}

\bsa \label{varhthetaest} 
Let $X_{ik} \sim F_i$, $\kg n_i$; $i=1,2$, be \ind\ and identically distributed \rvs, where $F_i(x) = \frac12 \left[ F_i^+(x) + F_i^-(x) \right]$ {\color{black} and assume that $(X_{1k})_{k=1}^{n_1}$ is \ind\ from $(X_{2\ell})_{\ell=1}^{n_2}$.} Further let $U_N= \sqrt{N} [(\olY_{2 \cdot} - \olY_{1 \cdot}) +1 - 2\theta]$, where $\olY_{i \cdot} = \frac1{n_i} \sumk {n_i} Y_{ik}$ and $Y_{1k} =F_2(X_{1k})$ and $Y_{2\ell} = F_1(X_{2 \ell})$ and let 
\bqan  
s_N^2 = \Var(U_N) &=& \frac{N}{n_1 n_2} \ (n_2 \sigma_1^2 + n_1 \sigma_2^2), \label{sNquadrat}
\eqan 
where the variances	$\sigma_1^2$ and $\sigma_2^2$ are defined in (\ref{sig1q}) and (\ref{sig2q}), respectively.
Further let $R_{ik}^* = R_{ik}-R_{ik}^{(i)}$ denote the placements in (\ref{r1kst}) and (\ref{r2lst}) and let $\olR_{i\cdot}^{\ *} = \frac1{n_i} \sumk {n_i} R_{ik}^*$, $i=1,2$ denote their means. Finally, let $\htau_N$ denote the unbiased estimator of $\tau$ in (\ref{tauNest}). Then the variance estimator
\bqan
\hsigma_N^2 &=& \frac1{(n_1-1)(n_2-1)} \left( \frac1{n_1n_2} (Q_1^2+Q_2^2) - \left[ \htheta(1-\htheta) - \tfrac14 \htau_N \right] \right)  \label{hsignub}
\eqan
has the following properties
\ben
  \item If $\sigma_1^2, \sigma_2^2 > 0$ and if $N/n_i \le N_0 < \infty$, $i=1,2$, then {\color{black} $N \hsigma_N^2$ is $L_2$-consistent for $s_N^2$ in (\ref{sNquadrat}) } in the sense that $E(N \hsigma_N^2/s_N^2 -1)^2 \to 0$.  \label{statement2b} 
\item For all samples sizes $n_1, n_2\ge2$ it holds that \\[-1.5ex]
 \ben 
  \item $E(\hsigma_N^2) = \sigma_N^2$,  \label{statement2a}  \\[-1.5ex]
  \item $0 \le \hsigma_N^2 \le \htheta(1-\htheta)/(m-1)$, where $m=\min\{ n_1, n_2\}$. Both these limits are sharp in the sense that there exist samples $X_{i1}, \ldots, X_{in_i}$, $i=1,2$, such that either $\hsigma_N^2 =0$, or $\hsigma_N^2 = \htheta (1-\htheta)/(m-1)$. \label{statement2b}
  \een 
\een
These results are true for data involving ties as well as for data without ties.
\esa
{\color{black} The derivation of Statement~\ref{statement2a} is given prior to Theorem~\ref{varhthetaest} while the proofs of the other statements are given in the Appendix.  }
\\[1ex]
\brems
\text{ } \\[-3ex]
\ben
\item Note that $\sqrt{N}(\htheta-\theta)$ is \asy ally equivalent to $U_N = \sqrt{N} \left[ (\olY_{2 \cdot} - \olY_{1 \cdot})  + 1 - 2 \theta \right]$. This means that the \asy\ variances of $\sqrt{N}\ \htheta$ and of $U_N$ are equal. For a formal proof, see, e.g., Brunner et al. (2019), Proposition~7.19, p.~386.
\item The result that $\hsigma_N^2 \le \htheta(1-\htheta)/(m-1)$ may be considered as an empirical version of the Birnbaum-Klose inequality (Birnbaum, 1956; Birnbaum and Klose, 1957). It is important, however, to note that this result does not state that it is valid for all estimators of $\sigma_N^2$ in (\ref{repsignq}). The statement is that it holds for the estimator $\hsigma_N^2$ in (\ref{hsignub}). Thus, this may be considered as a particular (noteworthy) property of this estimator. {\color{black} Moreover, it is easily seen that $\hsigma_N^2 \le 1 / [4(m-1)]$.}
\item For continuous \dbs\ it follows that $\tau=\htau_N=0$ and $\hsigma_N^2$ in (\ref{hsignub}) is equivalent to Hilgers' (1981) estimator using ranks or to Shirahata's (1993) estimator using indicator functions. In the case of ties, however, both these estimators are negatively biased if the ranks are simply replaced with the mid-ranks. Moreover, they may become negative in extreme cases. More details regarding comparisons with different variance estimators from the literature are given in the next Section.
\item {\color{black} The assumption $\sigma_1^2, \sigma_2^2 > 0$ implies that $s_N^2 > 0$ in (\ref{sNquadrat}) which is needed to derive the $L_2$-consistency of $\hsigma_N^2$.}
\een
\erems

\section{Comparison with Estimators from the Literature} \label{Biaslit}

There exist several estimators of $\sigma_N^2$ in the literature. Some of them are only valid in the case of no ties or are only asymptotically unbiased. 
The representation of $\sigma_N^2=Var(\htheta)$ in case of no ties is known since van Dantzig (1951) and in case of ties since Bamber (1975). 

{\color{black} Here we review some variance estimators from the literature and we discuss their properties.  Throughout this section, we will use the following notations for convenience:
\ben 
\item $d_N = n_1 (n_1-1) n_2 (n_2-1)$ \ and \ $m = \min\{n_1, n_2\}$
\item $\dl Q_i^2 = \sumk {n_i} \left( R_{ik}^* - \olR_{i \cdot}^* \right)^2$ as given in  (\ref{qi2qf}) \ and $Q_i^2/(n_i -1)$ are the empirical variances of the placements $R_{ik}^*, \ i=1,2$.
\item $\htheta$: Mann-Whitney estimator given in (\ref{f12plugin}) and (\ref{f12count})
\item $\htau_N$: estimator of the probability of ties in the overlap region of the \dbs\ as given in (\ref{tauNest}).
\een 

A brief overview of the following estimators is composed in Table~\ref{estimators} at the end of this section.

\subsection*{Sen-Hilgers-Shirahata Estimator}
 Estimators of $\sigma_N^2$ were already suggested by Sen (1967) and Govindarajulu (1968) assuming continuous \dfs. Later, Hilgers (1981) provided the estimator $\hsigma_{SHS}^2$ in Table~\ref{estimators} using ranks also assuming continuous \dfs\ and he showed that his estimator is identical to Sen's estimator corrected by $1/n_i^2$.  Halperin et al. (1987) also mentioned this typo in Sen's estimator in their paper. 

Neither Sen nor Hilgers discussed whether this estimator could become negative. This was briefly discussed by Shirahata (1993) who developed the same estimator based on indicator functions also assuming continuous \dfs. He discussed that this estimator, $\hsigma_U^2$ in his notation, could become negative mentioning that in his simulations, however, this did not happen. Since the estimators of Sen (1967), Hilgers (1981), and Shirahata (1993) are identical and are special cases of the estimator $\hsigma_N^2$ in (\ref{hsignub}), Theorem~\ref{varhthetaest}, Statement~\ref{statement2b}, shows that Shirahata's conjecture is incorrect and thus, $\hsigma_U^2\ge 0$. 

If, however, in the case of ties the ranks in Hilgers' estimator $\hsigma_{SHS}^2$ are simply replaced with mid-ranks, then the resulting estimator (without adding the term $n_1 n_2 \htau_N / 4$) is no longer unbiased and it can become negative as demonstrated by the following counter-example. 

Let $X_{1k} = \{1, 1, 2, 2, 3 \}$ and let $X_{2\ell} = \{3, 4, 4, 4, 5 \}$. Then $\hsigma_N^2 = 0.0004$ while $\hsigma_{SHS}^2 = -0.000225$. Note that $\hsigma_N^2$ in (\ref{hsignub}) contains the term $n_1 n_2 \htau_N / 4$. 

This shows that it is not trivial to provide an unbiased and non-negative estimator which is also valid in the case of ties. 

\subsection*{Cliff Estimator}
Cliff (1993) provides an unbiased estimator of $\sigma_N^2$ including the case of ties which is similar to Hilgers' estimator. But this estimator may become negative as discussed in his paper. He suggested to limit this estimator by $(1-\htheta^2)/(n_1n_2)$. But then he mentions that this substitution introduces a bias. Therefore we do not consider this estimator in more detail.

\subsection*{Bamber Estimator}
Allowing for ties, Bamber (1975) presented an estimator by using some involved explanations and without a formal proof stating that his estimator is unbiased. He did not discuss, however, whether this estimator may become negative. Also, formal proof of the consistency is not given. It turns out (for a formal derivation see Nowak et al., 2022) that Bamber's estimator is equivalent to $\hsigma_N^2$ in (\ref{hsignub}). This means that $\hsigma_N^2$ in (\ref{hsignub}) provides a convenient rank representation of Bamber's estimator.  

It may be noted that none of the subsequent papers by Halperin et al. (1987), Mee (1990), or Shirahata (1993) refers to Bambers's (1975) paper. Hanley and McNeil (1982) refer to Bamber's estimator but they assume no ties. It is quite astonishing that Bamber's unbiased estimator of the Mann-Whitney variance is not broadly perceived in the statistical literature. One reason may be {\color{black} that its representation is quite involved}. In this paper, we provide an unbiased estimator in (\ref{hsignub}) which is based on ranks, or more precisely on the placements $R_{1k}^*$ and $R_{2\ell}^*$ in (\ref{r1kst}) and (\ref{r2lst}) and thus, has a convenient representation.  

\subsection*{DeLong Estimator}
DeLong et al. (1988) refer to Bamber's estimator. But in their paper they develop an estimator of the \cm\ in a multivariate model which reduces to the variance estimator $\hsigma_{DL}^2$ in Table~\ref{estimators} in a univariate model.

This estimator is identical to the estimator $\hsigma_{BF}^2$ as derived by Brunner and Munzel (2000) in the nonparametric Behrens-Fisher situation. Brunner and Munzel showed that $\hsigma_{BF}^2$ (and in turn $\hsigma_{DL}^2$) is $L_2$-consistent for $s_N^2$ in the sense that $E( \hsigma_{BF}^2/s_N^2 -1)^2 \to 0$, if $N/n_i \le N_0 < \infty$, $i=1,2$. 

For small sample sizes, DeLong's estimator $\hsigma_{DL}^2$ in Table~\ref{estimators} may be biased. This can be easily seen from (\ref{EQ1qdetail}) and (\ref{EQ2qdetail}) and one obtains
\bqan
 E( \hsigma_{DL}^2) &=& \sigma_N^2 + \frac1{n_1n_2} \Big[ \underbrace{\theta(1-\theta) - (\sigma_1^2 + \sigma_2^2)}_{\ge 0} - \tfrac14 \tau \Big] \  . \label{biasDL}
\eqan
Then, from van Dantzig's inequality (1951), $\sigma_1^2 + \sigma_2^2 \le \theta(1-\theta)$, it follows that $\hsigma_{DL}^2$ may be unbiased or biased in both directions depending on ties and on whether 
$\theta(1-\theta) \neq \sigma_1^2 + \sigma_2^2$.

\subsection*{Perme-Manevski Estimator}
In a quite recent paper, Perme and Manevski (2019) state that the DeLong estimator $\hsigma_{DL}^2$ is not 'exact' and that DeLong et al. (1988) and Bamber (1975) propose an asymptotic non-parametric estimator. They do not mention, however, that Bamber (1975) stated that his estimator is unbiased for all sample sizes $n_i \ge 2$. Instead, they propose an alternative estimator $\hsigma_{PM}^2$ which they call 'exact'. This estimator is also listed in Table~\ref{estimators}. From (\ref{EQ1qdetail}) and (\ref{EQ2qdetail}) it follows that 
\bqa
E(\hsigma_{PM}^2) &=& \sigma_N^2 + \frac1{n_1n_2} \Big[ \underbrace{2[\theta(1-\theta) - (\sigma_1^2 + \sigma_2^2)]}_{\ge 0} - \tfrac14 \tau + O(\tfrac1N) \Big].
\eqa 

By van Dantzig's inequality it follows that in case of no ties, the bias may be larger than the one of the DeLong estimator since the non-negative expression $\theta(1-\theta) - (\sigma_1^2 + \sigma_2^2)$ is multiplied by 2. It may be unbiased or biased in both directions depending on ties and on whether 
$\theta(1-\theta) \neq \sigma_1^2 + \sigma_2^2$. Note that Perme and Manevski (2019) only consider the case of no ties. 

\subsection*{Hanley-McNeil Estimator}
Hanley and McNeil (1982) discuss two estimators of $\htheta$ in their paper. Regarding the first estimator, they refer to Bamber (1975). However, the estimator given in Eq.(1) in their paper is not Bamber's estimator. It is similar to $\hsigma_{PM}^2$ in Table~\ref{estimators} but using different weights for $Q_i^2$. Unfortunately, Bamber did neither discuss nor derive the properties of his estimator. When presenting their second estimator, Hanley and McNeil intended to provide a conservative estimator. They observed that in the case of exponential \dbs\ the variance of $\htheta$ was larger than for other \dbs. As in the case of exponential \dbs, the variance of $\htheta$ is a function only of $\theta$ and the sample sizes, they suggested an estimator $\hsigma_{HM}^2$ by plugging in $\htheta$ for $\theta$ in the variance representation for $\htheta$ assuming of exponential \dbs\
\bqa 
\Var_{exp}(\htheta \ ) &=& \frac1{n_1 n_2} \ \theta(1-\theta) \ \left[ 1 + (n_2-1) \frac{1-\theta}{2-\theta} + (n_1-1) \frac{\theta}{1+\theta} \right]
\eqa 
as given by Perme and Manevski (2019) who mention that 'there is no theoretical reason for consistency (of this estimator) in case of other distributions'. Moreover, the variance of $\htheta$ in the case of exponential \dbs\ is not the maximum possible variance of $\htheta$. In fact, $\sigma_{\max}^2 = \theta(1-\theta)/m$, as shown by the Birnbaum-Klose (1957). Because of these flaws, we will not consider $\hsigma_{HM}^2$ in more detail. It is somewhat unfortunate that later Newcombe (2006) recommended a modification of this estimator as a basis for the derivation of a \ci\ for $\theta$.

Some more estimators are considered in the literature for particular models where mainly continuous \dfs\ are assumed. As they are more or less similar tp the DeLong estimator and are restricted to to particular models, we do not consider them here in detail.

Obviously, there is an uncertainty in the statistical literature as to which variance estimator of $\htheta$ should be used.

\begin{table}[h]
{\color{black} 
\caption{List of some estimators of $\sigma_N^2 = \Var(\htheta)$ discussed in the literature. The estimators as functions of $d_n, m, \htheta, Q_1^2, Q_2^2, and \htau_N$ are listed in the left column, and some properties of the estimators in the right column.} \label{estimators}
\bcen 
\btb{lcl} \hline
& \hspace*{5ex} & \\[-1.5ex]
\mc{1}{c}{Estimator} & & \mc{1}{c}{Properties} \\[1ex] \hline 
& & \\[-1.5ex]
\bmp[h]{52ex} 
Sen = Hilgers = Shirahata \\[2ex] 
$\dl \hsigma_{SHS}^2 = \frac1{d_N} \left[Q_1^2 + Q_2^2 - n_1 n_2 \htheta(1-\htheta) \right]$ \\
\emp 
& \hspace*{5ex} & 
\bmp[h]{40ex}
only valid in case of no ties \\
unbiased \\ 
non-negative. \\[3ex]
\emp
\text{ } \\[4ex] \hline
& \hspace*{5ex} & \\[-1.5ex]
\bmp[h]{52ex} 
Bamber  \\[1ex]
$\dl \hsigma_N^2 = \frac1{d_N} \left( Q_1^2 + Q_2^2 - n_1 n_2\left[ \htheta(1-\htheta) - \tfrac14 \htau_N \right] \right)$ \\
\emp 
& \hspace*{5ex} &
\bmp[h]{40ex}
valid in general \\ 
unbiased \\
non-negative \\[1ex]
$\hsigma_N^2 \le \htheta(1-\htheta) / (m-1)$ \\[1ex]
\emp
\text{ } \\[3ex] \hline
& \hspace*{5ex} & \\[-1.5ex]
\bmp[h]{45ex} 
DeLong-DeLong-Ckarke-Pearson  \\[1ex]
$\dl \hsigma_{DL}^2 = \frac1{d_N} \left[ \left(1-\tfrac1{n_2} \right) Q_1^2 + \left(1-\tfrac1{n_1} \right) Q_2^2 \right]$ \\
\emp 
& & 
\bmp[h]{40ex}
may be unbiased  \\ 
or biased in both directions \\
depending on ties and whether \\[0.5ex] 
$\theta(1-\theta) > \sigma_1^2 + \sigma_2^2$ \\[1ex] 
\emp
\text{ } \\[3ex] \hline
& \hspace*{5ex} & \\[-1.5ex]
\bmp[h]{52ex} 
Perme-Manevski  \\[1ex]
$\dl \hsigma_{PM}^2 = \frac1{d_N} \bigg[\left(1-\tfrac1{n_2} \right)^2 Q_1^2 + \left(1-\tfrac1{n_1} \right)^2 Q_2^2$ \\
\hspace*{24ex} $ + (n_1-1) \ (n_2-1) \ \theta(1-\theta) \bigg] $ 
\emp
& & 
\bmp[h]{40ex}
may be unbiased  \\ 
or biased in both directions \\
depending on ties and whether \\[0.5ex] 
$\theta(1-\theta) > \sigma_1^2 + \sigma_2^2$ \\[1ex] 
\emp
\\[0.5ex] 
\text{ } \\ \hline
\etb 
\ecen 
}
\end{table}
}

{\color{black}
\section{Simulations} \label{simus}

\subsection{Simulated Biases of the Estimators $\hsigma_N^2$, $\hsigma_{DL}^2$, and $\hsigma_{PM}^2$} 
\label{simbias}

To demonstrate the amount of the different biases and the dependency on the underlying \dbs, a simulation study was performed where the following \dbs\ were selected:
\ben
\item Normal \dbs\ $N(0, \sigma_1^2), \ N(\delta, \sigma_2^2)$
\item $D_{\max}$-\dbs\ generating the maximal variance $\sigma_{\max}^2 = \theta(1-\theta)/m$ of $\htheta$, where, $\theta = \int F_1 dF_2$ and \\[1ex]
\btb{ll}
$F_1(x) \ = \ \left\{ \bay{ll} 0, & \ x \le 0, \\ \theta x, & \ 0 < x \le 1, \\ \theta, & \ 1 < x \le 2, \\ 
                               (1-\theta)x + 3\theta-2, & \ 2 < x \le 3, \\ 1, & \ x>3 \eay \right.$ 
& 
$F_2(x) \ = \ \left\{ \bay{ll} 0, & \ x \le 0, \\ x-1, & \ 1 < x \le 2, \\ 1, & \ x > 1 \ .  \eay \right.$
\etb
\text{ }  \\[0.5ex]
\item Poisson \dbs\ $X_i \sim Po(\lam_i), \ i=1,2$, with parameters $\lam_1=1$ and $\lam_2 = 1, \ldots, 13$
\item $5$-points ordinal scale \dbs\ for ordered categorical data which are generated by discretizing the \obss\ $X^B_i \sim \text{Beta}(x|a_i, b_i)$, $i=1,2$ from the Beta \dbs\ Beta$(x|a_i, b_i)$ such that $X^{ord}_i = \text{INT}(5 X^B_i|a_i, b_i) + 1$. Here we have selected the the Beta$(2,15)$-\db\ and the  Beta$(a_2,15)$-\db\ for $a_2=2, \ldots, 8$.
\een

Since all estimators $\hsigma_N^2, \hsigma_{DL}^2$, and $\hsigma_{PM}^2$ are asymptotically unbiased, we performed the simulations for the small sample sizes $n_1=n_2=10$ as an example. Figure~\ref{exabias} shows the biases of the above-mentioned variances estimators. 
The data were generated by the above listed \dfs, normal, $D_{\max}$, Poisson, and a $5$-points ordinal scale \dbs. These graphs demonstrate that in a practical data example, the actual type-I error $\alpha^*$ of a test and for a \ci\ the actual coverage probability $1-\alpha^*$ do not only depend on the quality of the approximation of $\htheta$ by a normal \db\ but also depend on an unknown bias of the variance estimator of $\htheta$. Thus, it appears advisable to remove this additional uncertainty by using an estimator that has only a small or even nor bias. 

It shall be noted here that these considerations only matter in case of small sample sizes where 'small' is a quite vague expression that depends on the underlying \dfs\ and the effect $\theta$. For sample sizes $n_1=n_2=50$, for example, the ratio $\hsigma_{PM}^2/\sigma_N^2 \ge 1.1$ for $\theta \ge 0.98$ in case of normal \dbs\ or ordinal data on a $5-$points scale. Sample sizes of about $n_1=n_2=500$ are necessary to obtain a similar graph for the same ratio as for $n_1=n_2=10$ in the case of the $D_{\max}$-\dbs\ in Fig.~\ref{exabias} (upper row, right). 
For discrete \dbs, e.g., in case of a $5$-points ordinal scale, also for quite a small effect close to $\theta = \nfrac12$, it is possible that the ratio $\hsigma_{PM}^2/\sigma_N^2 \approx\ 1.3$ for $n_1 = n_2 = 10$ while a samples size of about 100 per group reduces this bias to $\approx\ 1.02$. 

This motivates using the unbiased estimator $\hsigma_N^2$ in (\ref{hsignub}) first derived by Bamber (1975) and studied in more detail in this paper.

\bfg[h] 
{\color{black}
\includegraphics[width=41ex,height=32ex]{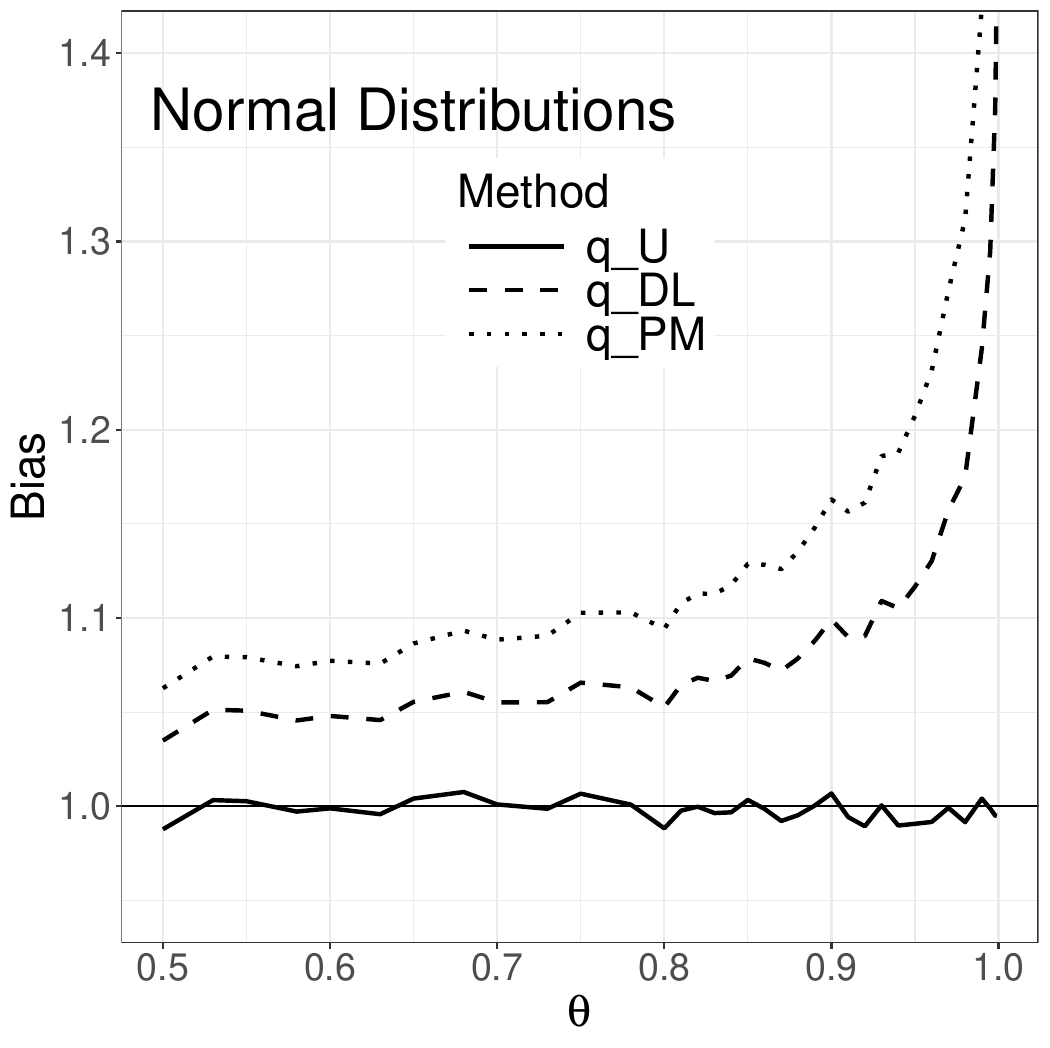} \sep \
\includegraphics[width=41ex,height=32ex]{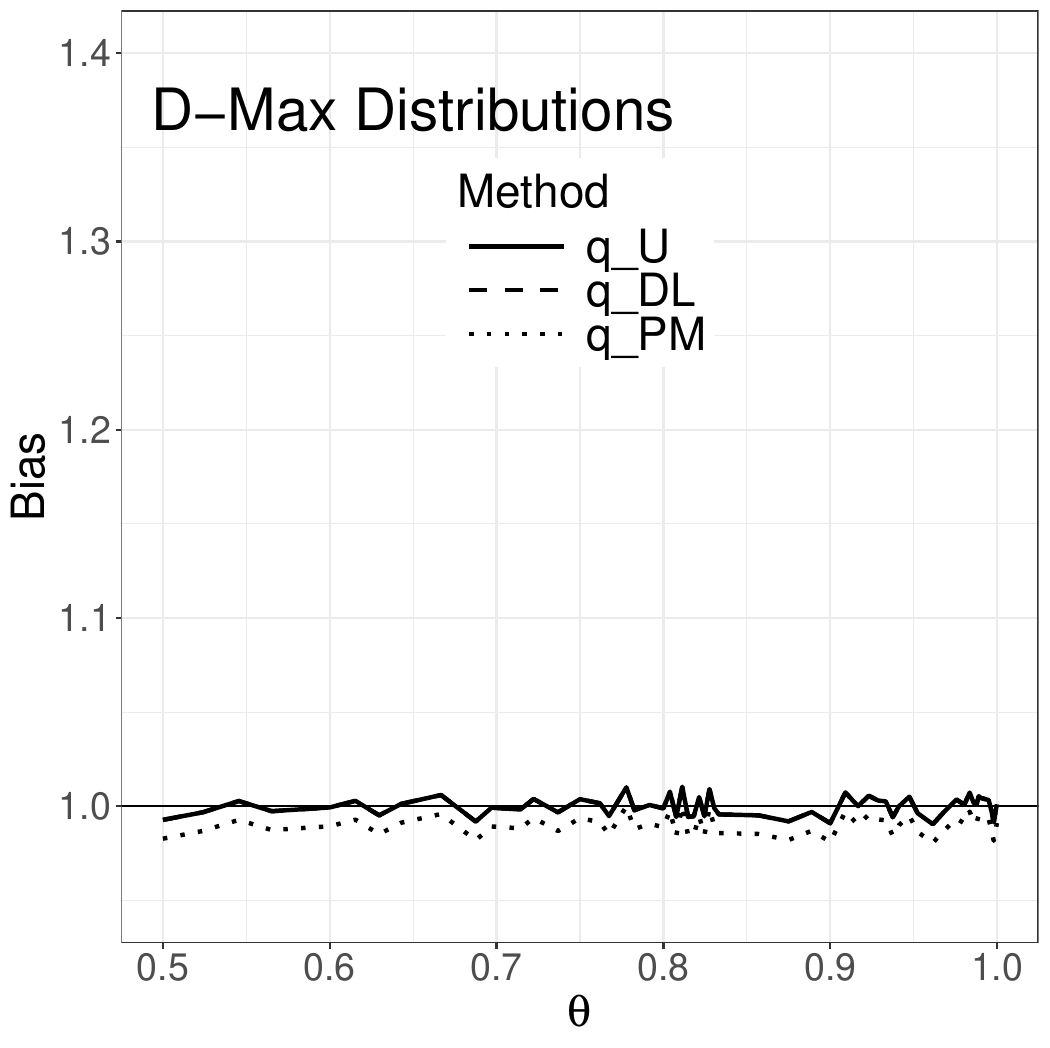} \\[1ex]
\includegraphics[width=41ex,height=32ex]{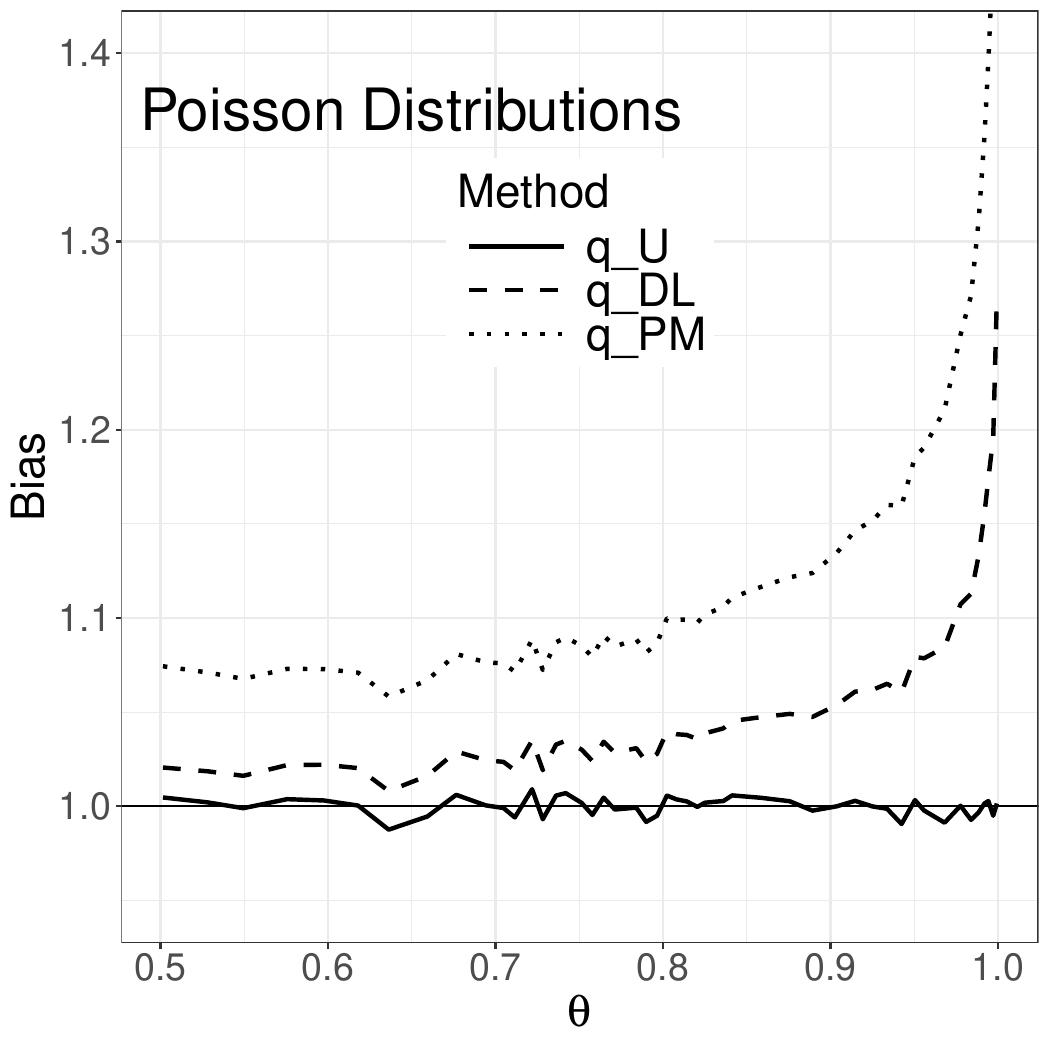} \sep \ 
\includegraphics[width=41ex,height=32ex]{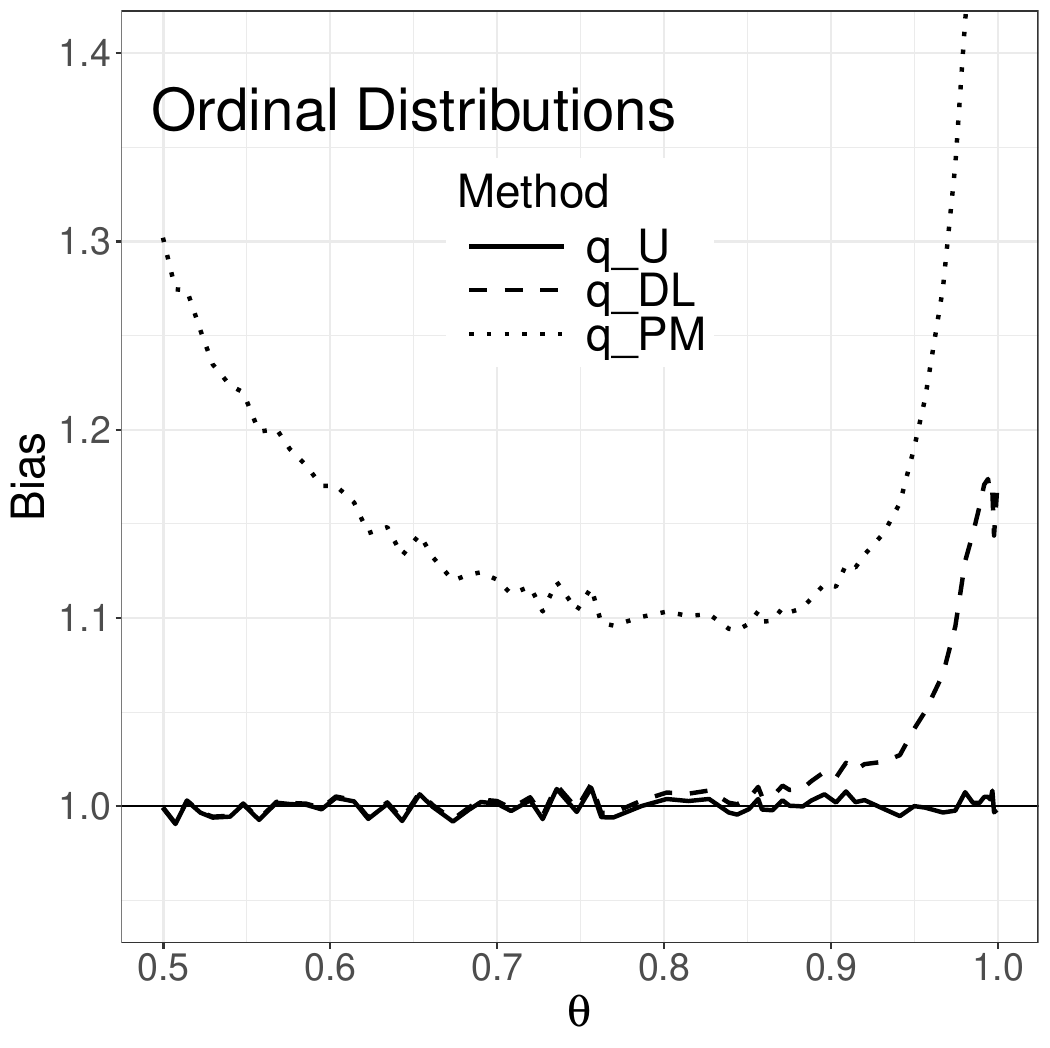} \\
\caption{Biases of the variance estimators $\hsigma_N^2$ (solid), $\hsigma_{DL}^2$ (dashed), and 
$\hsigma_{PM}^2$ (dotted) displayed in Table~\ref{estimators} for the samples sizes $n_1=n_2=10$. 
The data were generated from normal \dbs\ (upper row, left), $D_{\max}$-\dbs\ (upper row, right), 
Poisson-\dbs\ (lower row, left), and a 5-points ordinal scale (lower row, right).
} \label{exabias}
}
\efg  

\subsection{MSE of the Estimators $\hsigma_N^2$, $\hsigma_{DL}^2$, and $\hsigma_{PM}^2$} \label{msesim}

A summary measure for the quality of an estimator which includes the bias as well as the variance of the estimator is the so-called 'means-squared-error' (MSE),
\bqa 
\text{MSE}(\hlam, \lam) &=& \Var(\hlam) + \left[E(\hlam) - \lam \right]^2,
\eqa 
where $\lam$ denotes the parameter to be estimated, $\hlam$ an estimator of $\lam$, and $\var(\hlam)$ the variance of this estimator. As seen from Fig.~\ref{exabias} the biases of the estimators $\hsigma_N^2$, $\hsigma_{DL}^2$, and $\hsigma_{PM}^2$ depend on $\theta$. Thus, we consider these biases relative to that quantity which it should estimate. Here, the variance $\sigma_N^2$ in (\ref{repsignq}) shall be estimated and the relative MSEs are given by
\bqa
q-\text{MSE}(\hsigma_N^2, \sigma_N^2) &=& \frac1{\sigma_N^2} \left(\Var(\hsigma_N^2) + \left[ E(\hsigma_N^2) - \sigma_N^2 \right]^2   \right) \ = \ \frac1{\sigma_N^2} \Var(\hsigma_N^2) \\
q-\text{MSE}(\hsigma_{DL}^2, \sigma_N^2) &=& \frac1{\sigma_N^2} \left(\Var(\hsigma_{DL}^2) + \left[ E(\hsigma_{DL}^2) - \sigma_N^2 \right]^2   \right) \\
q-\text{MSE}(\hsigma_{PM}^2, \sigma_N^2) &=& \frac1{\sigma_N^2} \left(\Var(\hsigma_{PM}^2) + \left[ E(\hsigma_{PM}^2) - \sigma_N^2 \right]^2   \right) \ .
\eqa 

Basically, $\sigma_N^2$ is a scaling factor. It turned out in our simulations that $\hsigma_N^2$ has the smallest q-MSE in all cases. As examples, we display the results for normal \dbs, $5$-points ordinal scale \dbs, exponential \dbs, and $D_{\max}$-\dbs\ in Figure~\ref{qMSE}.

\bfg[h] 
{\color{black}
\hspace*{3ex}
\includegraphics[width=38ex,height=25ex]{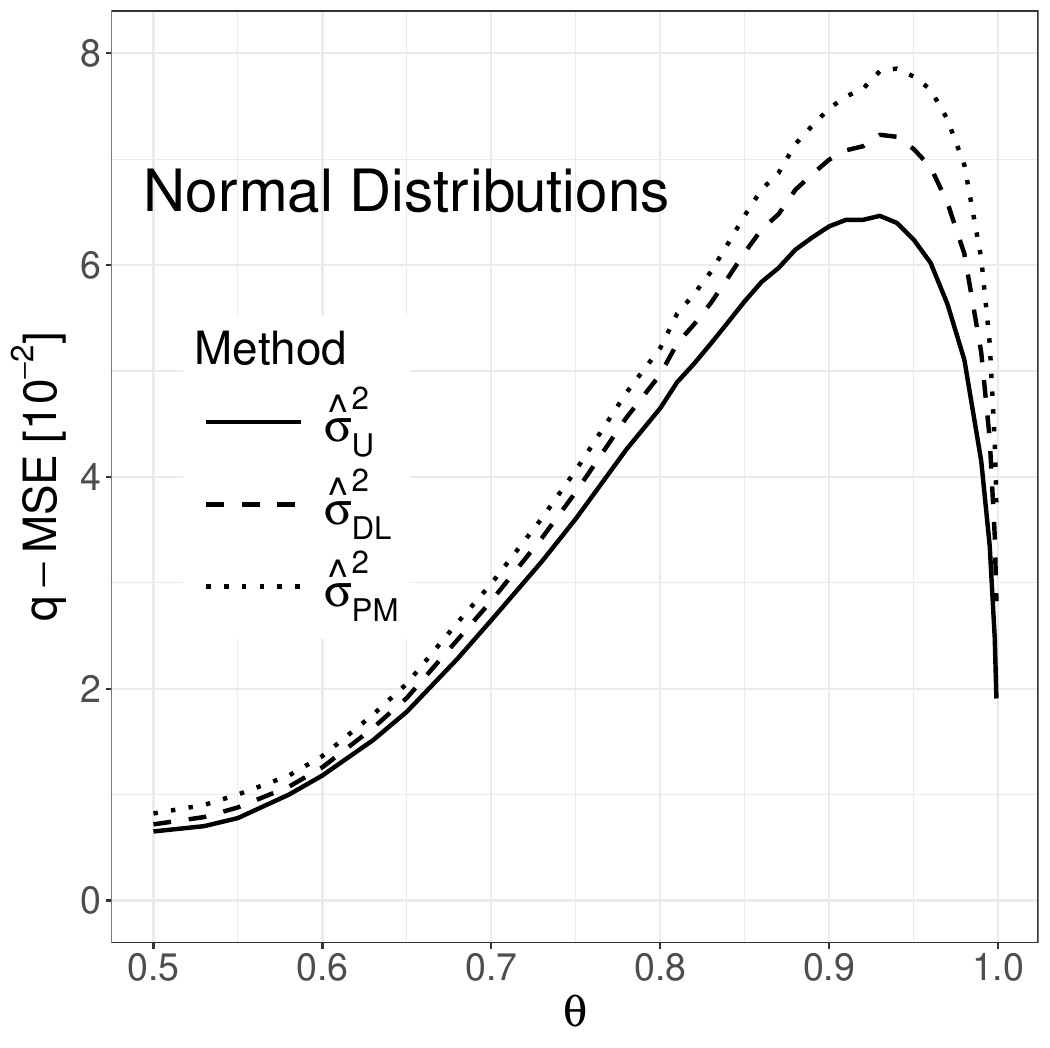} \sep \
\includegraphics[width=38ex,height=25ex]{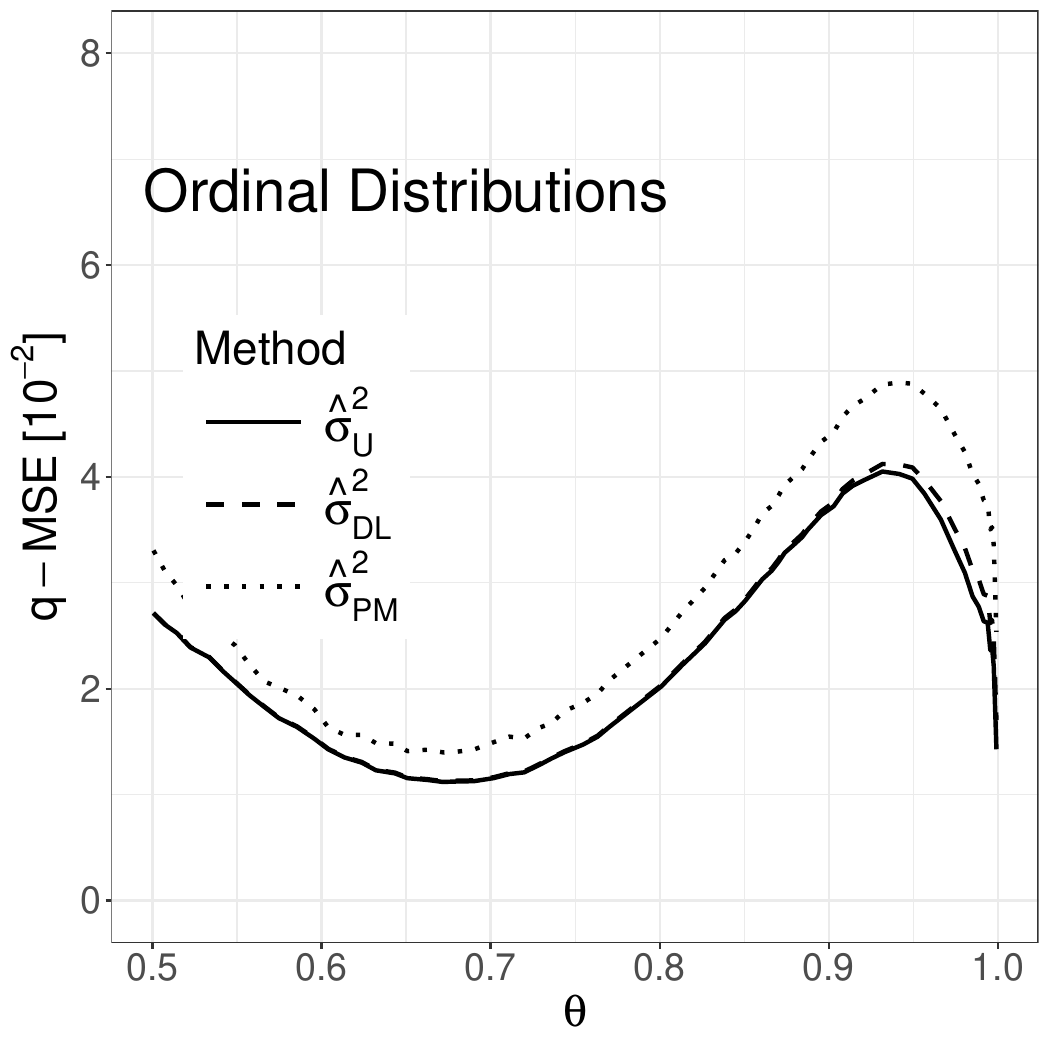} \\[1ex]
\hspace*{3ex} \includegraphics[width=38ex,height=25ex]{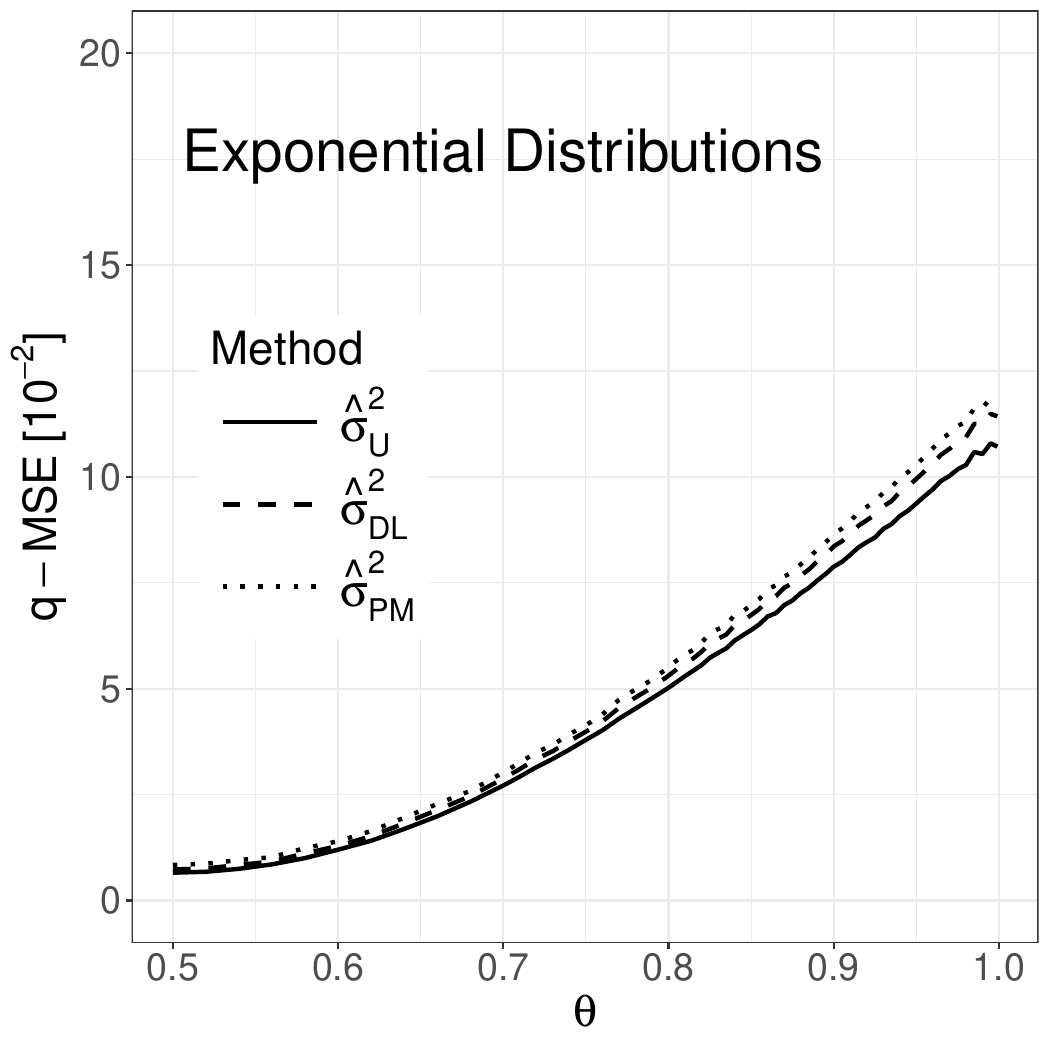} \sep \ 
\includegraphics[width=38ex,height=25ex]{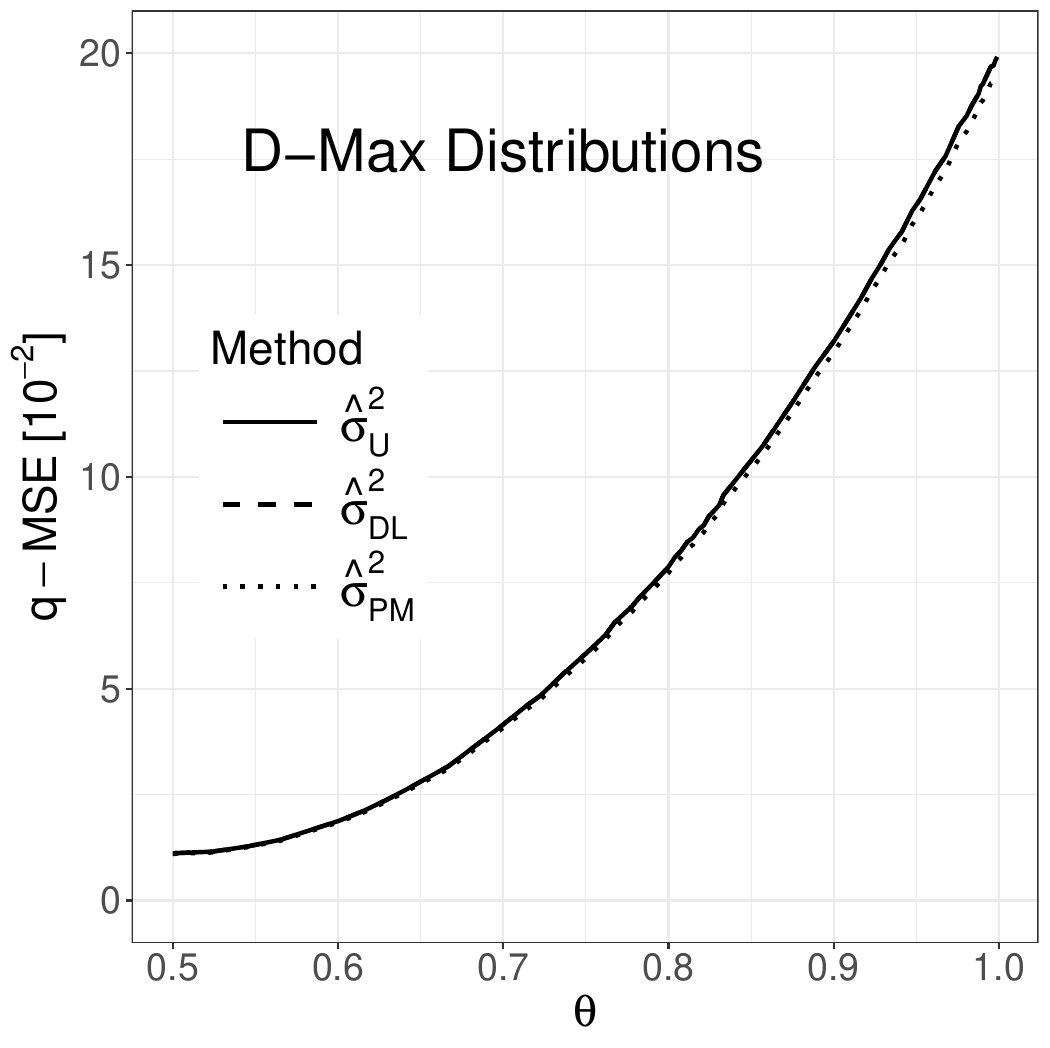} \\
\caption{Relative q-MSEs of the estimators $\hsigma_N^2$ (solid), $\hsigma_{DL}^2$ (dashed), and 
$\hsigma_{PM}^2$ (dotted) in $[10^5]$ simulations runs for each setting for the sample sizes $n_1=n_2=10$. 
The alternatives $\theta \in [0.5, 0.999]$ are generated from normal \dbs\ (upper row, left), $5$-points ordinal scale \dbs\ (upper row, right), exponential \dbs\ (lower row, left), and $D_{\max}$-\dbs\ (lower row, right). Note that the scales of the ordinate are different in the upper and lower row to better demonstrate the differences between the estimators.} \label{qMSE}
\text{ } \\
The conclusion from this simulation study is that the unbiased estimator $\hsigma_N^2$ (Bamnber's estimator) can be recommended as a 'good variance estimator' for the Mann-Whitney variance. It outperforms the DeLong- and Perme-Manevski estimators.

}
\efg  

}

\section{Discussion}\label{disc}

We intended to derive an unbiased variance estimator of the well-known Mann-Whitney variance $\sigma_N^2$ in (\ref{repsignq}). In the case of continuous \dfs\ $F_1(x)$ and $F_2(x)$, the variance representation is known since van Dantzig (1951). But it seems that Bamber (1975) was the first to provide a representation of $\sigma_N^2$ which is also valid in the case of ties. The representation of Bamber's estimator, however, is quite involved.  Moreover, the basic properties - except unbiasedness - of this estimator are not discussed in Bamber's paper. In particular, it is necessary that an unbiased variance estimator cannot become negative since in {\color{black} many} applications the standard deviation of this estimator is used. If the estimator is set equal to 0 if it is negative, then it is no longer unbiased.

To the best of our knowledge, it has not been shown for any unbiased variance estimator of the Mann-Whitney variance which is also valid for ties that it cannot become negative. Bamber (1975) does not discuss this topic and Cliff (1993) states that the estimator given in Equation (9) in his paper may become negative. 
Neither Sen (1967) nor Hilgers (1981) discuss whether their estimators can become negative. Moreover, these estimators are only valid in case of no ties. The proposed estimator $\hsigma_N^2$ in (\ref{hsignub}), however, is also valid for ties, and it is shown in the appendix that $\hsigma_N^2 \ge 0$. In addition, the \cms\ of the placements $R_{1k}^*$ in (\ref{r1kst}) and $R_{2\ell}^*$ in  (\ref{r2lst}) are derived. Since many variance estimators provided in the literature are based on the quadratic forms $Q_i^2$ in (\ref{qi2qf}) of the centered placements, this derivation enables a simple computation of the bias of the different estimators. 

{\color{black} 
The key point in the derivation of $\hsigma_N^2$ is that the \cms\ of the placement vectors $\vR_1^*$ and 
$\vR_2^*$ in (\ref{csvi})  have compound symmetry structures. The expectations of the quadratic forms $Q_i^2$ in (\ref{qi2qf}) based on $\vR_1^*$ and $\vR_2^*$ are not unbiased estimators of $s_1^2 = \Var(R_{11}^*)$ or $s_2^2 = \Var(R_{21}^*)$. Computing the bias by Lancaster's theorem and then appropriately estimating the bias, leads to the unbiased estimator $\hsigma_N^2$.

Finally, the behavior of $\hsigma_N^2$ compared with competing estimators, the mean-squared-error (MSE) is  commonly used. Note that the variance $\sigma_N^2$ depends on $\theta$. A 'uniform' superiority of an estimator, however, requires that the MSE of this estimator should be uniformly larger than that of a competing estimator. Then graphing the MSE over $\theta$ scaled by $\sigma_N^2$ can show whether this estimator is uniformly preferable. 

The simulation study in Sect.~\ref{msesim} shows that $\hsigma_N^2$ is preferable to $\hsigma_{DL}^2$ and $\hsigma_{PM}^2$ for some important classes of \dfs\ such as a normal \db, exponential \db, and $5$-points ordinal \dbs. For the extreme case of $D_{\max}$-\dbs\ it turns out that $\hsigma_N^2$, $\hsigma_{DL}^2$, and $\hsigma_{PM}^2$ are comparable. We have also investigated some more \dbs, such as Laplace-, $\gamma$-, and Beta-\dbs\ (not shown here). Also in these examples it turns out that $\hsigma_N^2$ is always superior to $\hsigma_{DL}^2$ and $\hsigma_{PM}^2$ regarding the scaled MSE. 

In this paper, Bamber's estimator $\hsigma_N^2$ of the Mann-Whitney variance is studied in detail. On the other hand, a convenient representation using some simple rankings is given. On the other hand, it is shown that this estimator cannot become negative (disproving Shirahata's conjecture). Moreover, a sharp upper bound of $\hsigma_N^2$ is derived which can be regarded as an empirical version of the well-known Birnbaum-Klose inequality which is a noteworthy property of this estimator.
Finally, it is demonstrated in a simulation study that this estimator outperforms the commonly used  variance estimators $\hsigma_{DL}^2$ and $\hsigma_{PM}^2$ by a uniformly smaller MSE relative to $\sigma_N^2$ as graphically shown in Fig.~\ref{qMSE}. 

Unfortunately, Bamber's estimator $\hsigma_N^2$ is not well perceived in the statistical literature and we hope that the properties of it derived in this paper will contribute to its larger popularity and a more frequent use. It is easy to compute and has an intuitive representation using rankings, it is unbiased and cannot become negative, and finally outperforms commonly used estimators by a smaller MSE.
}

\section{Appendix}

\section{Proof of Theorem~\ref{varhthetaest}} \label{sigmahatproof}

\subsection{$L_2$-Consistency of the Variance Estimator $\hsigma_N^2$} \label{l2cons}

To show $L_2$-consistency of $\hsigma_N^2$ it suffices to show that $E\left[ N \hsigma_N^2 - s_N^2 \right]^2 \to 0$ since $s_N^2 > 0$ by the assumption that $\sigma_1^2, \sigma_2^2 >0$. Note that $s_N^2 = N(\sigma_1^2/n_1 + \sigma_2^2/n_2)$ and let
\bqa
\hsigma_1^2 = \frac{Q_1^2}{n_2^2(n_1-1)} , & \text{and} & \hsigma_2^2 = \frac{Q_2^2}{n_1^2(n_2-1)} \ .
\eqa 

Then straightforward computations show that 
\bqa 
N \left[\hsigma_N^2 - \left( \tfrac{\sigma_1^2}{n_1} + \tfrac{\sigma_2^2}{2_1} \right) \right] &=& 
N \Bigg[ \frac{n_2}{n_1} \left( \frac{\hsigma_1^2}{n_2-1} - \frac{\sigma_1^2}{n_2} \right) + \frac{n_1}{n_2} \left( \frac{\hsigma_2^2}{n_1-1} - \frac{\sigma_2^2}{n_1} \right) \\
  & & \hspace*{20ex} - \ \frac1{(n_1-1)(n_2-1)} \left[ \htheta(1-\htheta) - \tfrac14 \htau_N \right] \Bigg] 
  \ .
\eqa 

Now note that $0 < \sigma_i^2 \le 1$, $0 \le \hsigma_i^2 \le 1$, $i=1,2$, and that $0 \le \htheta(1-\htheta) \le \nfrac14$ and $0 \le \htau_N \le1$. Then, using Jensen's inequality and taking expectations it follows from the assumption $N/n_i \le N_0$, that
\bqa 
E \left( N \left[\hsigma_N^2 - \left( \tfrac{\sigma_1^2}{n_1} + \tfrac{\sigma_2^2}{2_1} \right) \right] \right)^2 & \le & 6N_0^2 E(\hsigma_1^2 - \sigma_1^2)^2 + O\left( \tfrac{N_0^2}{n_2^2} \right) \\ 
 & &  + 6 N_0^2 E(\hsigma_2^2 - \sigma_2^2)^2 + O\left( \tfrac{N_0^2}{n_1^2} \right) + O\left( \tfrac{N_0^2}{n_1 n_2} \right) \ .
\eqa 

Then the result follows by noting that $E(\hsigma_i^2 - \sigma_i^2)^2 \to 0$ for $N \to \infty$, $i=1,2$, such that  $N/n_i \le N_0$ and if $\sigma_1^2, \sigma_2^2 >0$ (for details see Brunner and Munzel, 2000).



\subsection{Non-Negativity of the Variance Estimator $\hsigma_N^2$} \label{hsigmanonneg}

It is not straightforward to show that $\hsigma_N^2 \ge 0$ using its rank representation in 
(\ref{hsignub}). Therefore, we re-write $\hsigma_N^2$ as sums of squares and products of count functions
which, of course, are all non-negative as defined in Sect.~\ref{MWplace}. Let 
\begin{align}
    \mathcal{A} &= \sum_{k=1}^{n_2}\sum_{r=1}^{n_1} c(X_{1r},X_{2k})^2, 
    &\mathcal{B} = \sum_{k=1}^{n_2}\sum_{r\not=r'}c(X_{1r},X_{2k})c(X_{1r'},X_{2k}), \nonumber\\
    \mathcal{C} &= \sum_{k\not =k'}\sum_{r=1}^{n_1}c(X_{1r},X_{2k})c(X_{1r},X_{2k'}),
    &\mathcal{D} = \sum_{k\not =k'}\sum_{r\not= r'} c(X_{1r},X_{2k})c(X_{1r'},X_{2k'}),\label{ABCD}
\end{align}
and for convenience let $\kS = \kA + \kB + \kC + \kD$ and note that $\kS \ge 0$. Further, note that 
$R_{2k}^\ast=\sum_{r=1}^{n_1}c(X_{1r},X_{2k})$ by (\ref{r2lst}), $R_{1r}^\ast = \sum_{k=1}^{n_2} c(X_{2k},X_{1r})$ by (\ref{r1kst}), and 
\bqan
 \htheta^2 &=& \frac1{n_1^2 n_2^2} \sumr {n_1} \sum_{r'=1}^{n_1} \sumk {n_2} \sum_{k'=1}^{n_2} 
               c(X_{1r}, X_{2k}) c(X_{1r'}, X_{2k'}) \ = \ \frac1{n_1^2 n_2^2} \kS \label{thdq} 
\eqan
by (\ref{f12count}). 
Then, 
\bqan
Q_2^2 &=& \sum_{k=1}^{n_2}\left(R_{2k}^\ast - \overline{R}_{2\cdot}^\ast\right)^2 \ = \ 
          \sumk {n_2} \left( R_{2k}^\ast \right)^2 - n_2 \left( \olR_{2\cdot}^\ast \right)^2 \nnr\ \\
      &=&\sumk {n_2}\sum_{r\not=r'}^{n_1} c(X_{1r},X_{2k}) c(X_{1r'},X_{2k}) + \sumk {n_2} \sumr {n_1} 
			   c(X_{1r},X_{2k})^2 - n_1^2n_2 \htheta^2 \label{q2q} \nnr\ \\
      &=& \kA + \kB - \frac{1}{n_2} \kS \label{q2q} ,
\eqan       
by (\ref{thdq}) and noting that $\left( \olR_{2\cdot}^\ast \right)^2 =  n_1^2n_2 \htheta^2$. Furthermore, by using the relation $c(X_{2k}, X_{1r}) = 1 - c(X_{1r}, X_{2k})$, it follows in a similar way that 
\bqan       
Q_1^2 &=& \sumr {n_1} \left(R_{1r}^\ast-\olR_{1\cdot}\right)^2 \ = \ \sum_{r=1}^{n_1}\sum_{k=1}^{n_2
         }\sum_{k'=1}^{n_2} c(X_{2k},X_{1r}) c(X_{2k'},X_{1r}) - n_1 n_2^2(1-\htheta)^2 \nnr\ \\
		  &=& \sumr {n_1}\sumk {n_2}\sum_{k'=1}^{n_2} c(X_{1r},X_{2k}) c(X_{1r'},X_{2k'}) - n_1 n_2^2 
			    \htheta^2 \nnr \\
      &=& \sumr {n_1} \sum_{k\not=k'} c(X_{1r},X_{2k}) c(X_{1r'},X_{2k'}) + \sumr {n_1} \sumk {n_2} 
			     c(X_{1r},X_{2k})^2 - n_1 n_2^2 \htheta^2  \nnr\ \\
      &=& \kC + \kA - \frac1{n_1} \kS \ . \label{q1q} 
\eqan

Finally let $Q_3^2 = n_1 n_2 \left[(\htheta (1-\htheta) - \frac14 \htau_N \right]$. Then, by noting that 
$c^2(X_{1r}, X_{2k}) = c(X_{1r}, X_{2k}) - \frac14 \htau_N$ by (\ref{cqdef}), the quantity $Q_3^2$ can be written as
\bqan
Q_3^2 &=& \sumk {n_2} \sumr {n_1} \left(c(X_{1r},X_{2k}) -\htheta \right)^2 \ = \ \sumk {n_2} \sumr {n_1} c^2(X_{1r},X_{2k}) - n_1n_2 \htheta^2 \nnr\ \\
			&=& \kA - \frac1{n_1 n_2} \kS  \label{q3q}
\eqan
by using (\ref{thdq}).

Now let $\wtN = n_1(n_1-1) n_2(n_2-1)$. Then $\wtN \hsigma_N^2 = Q_1^2 + Q_2^2 - Q_3^2$ and by 
combining the results in (\ref{ABCD}), (\ref{thdq}), (\ref{q1q}), (\ref{q2q}), and (\ref{q3q}) it 
follows that 
\bqan
 \wtN \hsigma_N^2 &=& \kA + \kB - \frac1{n_2}\kS + \kA + \kC - \frac1{n_1}\kS - \kA + 
                      \frac1{n_1 n_2} \kS \nonumber\\
				   &=& \kS - \kD - \left(\frac1{n_1} + \frac1{n_2} - \frac1{n_1 n_2} \right) \kS , \; \text{respectively} \ \nonumber\\
       \hsigma_N^2 &=& \frac{1}{n_1^2n_2^2}\kS - \frac{1}{\wtN}\kD = \widehat{\theta}^2 - \frac{1}{\wtN}\kD \nonumber\\
&=& \frac{1}{n_1^2n_2^2} \left(\mathcal{A} + \mathcal{B} + \mathcal{C} -\frac{N-1}{(n_1-1)(n_2-1)}\mathcal{D}  \right). \label{sigqge0}
\eqan
Hence, the non-negativity of $\hsigma_N^2$ follows from the relationship between the sums of two count functions $\kA, \kB, \kC$ and $\kD$, respectively. For convenience, we define the additional sums of count functions
\begin{align} \label{EF}
    \kE = \sum_{k=1}^{n_2}\sum_{r=1}^{n_1}c(X_{1r},X_{2k})\;\;\; \text{and}  \;\;\;\kF=\sum_{k=1}^{n_2}\sum_{r=1}^{n_1} [ c^+(X_{1r},X_{2k}) - c^-(X_{1r},X_{2k})],
\end{align}
and note that $n_1n_2\widehat{\theta}=\kE$ and $n_1n_2\widehat{\tau}_N= \kF$. By (\ref{thdq}) it follows that
\bqa 
   \mathcal{A} +\mathcal{B}+ \mathcal{C} = n_1^2n_2^2\widehat{\theta}^2 -\mathcal{D}.
\eqa
Equivalently, it follows from \eqref{EF}, that  
\bqa
n_1^2n_2^2 \widehat{\theta}^2 = \kE^2 = \left(\kA + \tfrac14\kF\right)^2
\eqa
and therefore, 
\bqan \label{ABCAD}
\mathcal{A} +\mathcal{B}+ \mathcal{C} = \left(\kA + \tfrac14\kF\right)^2 -\mathcal{D}.
\eqan
The latter implies $\kD {\color{black} \le } \left(\kA + \tfrac14\kF\right)^2 = \kA+\kB+\kC+\kD$. Indeed, if $\mathcal{D}>0$, then there exist at least two pairs $(X_{1r}, X_{2k})$ and $(X_{1r'}, X_{2k'})$, $r\not = r'$  and $k\not = k'$, such that $c(X_{1r},X_{2k})\cdot c(X_{1r'},X_{2k'}) >0$. This implies, however, that at least three pairs $c(X_{1r},X_{2k}) >0, c(X_{1r'},X_{2k'})>0$, and either $c(X_{1r},X_{2k'})>0$ or  $c(X_{1r'},X_{2k})>0$ exist. Hence, whenever $\mathcal{D}>0$, then $\mathcal{A} >0$ and $\mathcal{B}>0$ or $\mathcal{C}>0$ (or both).  In particular, there must exist at least one pair $c(X_{1r},X_{2k})=1$, otherwise $\mathcal{D}$ would be 0. Since there are $n_1n_2$ terms in $\kA$, $n_2n_1(n_1-1)$ in $\kB$, $n_1n_2(n_2-1)$ in $\kC$ and $n_1(n_1-1)n_2(n_2-1)$ terms in $\kD$, it follows from \eqref{ABCAD}
\bqan 
\mathcal{A} +\mathcal{B}+ \mathcal{C} & \geq & \frac{n_1n_2 + n_1n_2(n_2-1) + n_2n_1(n_1-1)}{n_1(n_1-1)n_2 (n_2-1)} \kD, \nnr\ \\
& = & \frac{N-1}{(n_1-1)(n_2-1)}\mathcal{D} \ .\label{ABC}
\eqan 

Hence, plugging-in \eqref{ABC} into \eqref{sigqge0}$,\, \hsigma_N^2 \geq 0$. Note that $\hsigma_N^2 =0$, if $\widehat{\theta}=0$, since $\kA=\kB=\kC=\kD=0$. If $\widehat{\theta} = 1$, then $\mathcal{D}=n_1(n_1-1)n_2(n_2-1)$, $\mathcal{A}=n_1n_2$, $\mathcal{B}=n_2n_1(n_1-1)$ and $\mathcal{C}=n_1n_2(n_2-1)$ and hence $\widehat{\sigma}_N^2 = \frac{1}{n_1^2n_2^2} \left(\mathcal{A} + \mathcal{B} + \mathcal{C} -\frac{N-1}{(n_1-1)(n_2-1)}\mathcal{D}  \right)=0$.

\subsection{Verification of the upper bound}
The computation of the upper bound $\tfrac{\widehat{\theta}(1-\widehat{\theta})}{min(n_1,n_2)-1}$ of $\hsigma_N^2$ is \textemdash as the verification of the non-negativity of $\hsigma_N^2$ \textemdash a very challenging task using its rank-based version. We therefore use its representation with the sums of products of the count functions $\kA, \kB, \kC$, and $\kD$, given in \eqref{ABCD}, respectively. In particular, we use the representation 
\bqa
\hsigma_N^2 = \widehat{\theta}^2 - \frac{1}{\wtN}\kD
\eqa
as given in the line above \eqref{sigqge0}. In the following, let $m=\min(n_1,n_2)-1$ and let w.l.o.g. be $n_1\leq n_2$. The upper bound can now be verified in a few steps: 
\begin{eqnarray*}
&&\frac{\widehat{\theta}(1-\widehat{\theta})}{m} - \widehat{\sigma}_N^2 = \frac{\widehat{\theta}(1-\widehat{\theta})}{m} - \widehat{\theta}^2 + \frac{1}{\wtN}\kD
= \frac{1}{m}\widehat{\theta} - \frac{m+1}{m} \widehat{\theta}^2 + \frac{1}{\wtN}\kD \\
&=& \frac{1}{n_1-1}\widehat{\theta} -  \frac{n_1}{n_1-1} \widehat{\theta}^2 + \frac{1}{\wtN}\kD
 \ = \ \frac{1}{n_1-1} \left(\widehat{\theta} - n_1 \widehat{\theta}^2\right) + \frac{1}{n_1(n_1-1)n_2(n_2-1)} \mathcal{D} \\
&=& \frac{1}{n_1-1} \Biggl[ \frac{1}{n_1n_2}\underbrace{\sum_{k=1}^{n_2} \sum_{r=1}^{n_1} c(X_{1r},X_{2k})}_{\kE}-n_1\widehat{\theta}^2 \Biggr] + \frac{1}{n_1(n_1-1 )n_2(n_2-1)}\mathcal{D} \\
&=& \frac{1}{(n_1-1)n_1n_2} \Biggl[\underbrace{ \mathcal{E} - \frac{1}{n_2} \left(\mathcal{A}+\mathcal{B}+\mathcal{C}+\mathcal{D} \right) + \frac{1}{(n_2-1)}\mathcal{D}}_{\geq 0} \Biggr] \geq 0
\end{eqnarray*}


\newpage

\section{References}
\bdes

\item{Bam1975}
Bamber D (1975) The Area above the Ordinal Dominance Graph and the Area below the Receiver Operating Characteristic Graph. J Math Psychol 12: 387--415.

\item{Birnbaum1956}
Birnbaum ZW (1956) On a use of the Mann-Whitney statistic. In J Neyman (Ed) Proceedings of the Third Berkeley Symposium on Mathematical Statistics (pp. 13-17). Berkeley, Los Angeles: University of California Press.
   
\item{BiKl1957}
Birnbaum ZW, Klose OM (1957) Bounds for the variance of the Mann-Whitney statistic. Ann Math Stat 28: 933--945.

\item{BBK2019}
Brunner E, Bathke AC, Konietschke  F (2019) Rank and pseudo-rank procedures for independent observations in factorial designs. Springer Series in Statistics, Springer Heidelberg.

\item{BM2000}
Brunner E, Munzel U (2000) The nonparametric Behrens-Fisher problem: Asymptotic theory and a small - sample approximation. Biom J 42: 17--25.

\item{BruPu2001}
Brunner, E., Puri, M. L. (2001). Nonparametric methods in factorial designs. Statistical Papers, 42, 1-52.

\item{BruPu2002}
Brunner, E., Puri, M. L. (2002). A class of rank-score tests in factorial designs. Journal of Statistical Planning and Inference 103, 331-360.

\item{Buyse2010}
Buyse M (2010) Generalized pairwise comparisons of prioritized outcomes in the two-sample problem. Stat Med 29: 3245--3257.



\item{DLDLCP1988}
DeLong ER, DeLong DM, Clarke-Pearson DL (1988) Comparing the areas under two or more correlated receiver operating characteristic curves: a nonparametric approach. Biometrics 44: 837--845.


\item{Cliff1993}
Cliff  N (1993) Dominance statistics: Ordinal analyses to answer ordinal questions. Psychological Bulletin  114: 494--509.

\item{GFBBK2021)}
Gasparyan SB, Folkvaljon F, Bengtsson O, Buenconsejo J, Koch GG (2021) Adjusted win ratio with stratification: calculation methods and interpretation.  Stat Methods Med Res 30: 580--611.


\item{Gov1968}
Govindarajulu Z (1968) Distribution-free confidence bounds for Pr$\{X < Y \}$. Ann Inst Stat Math   20: 229--238.

\item{HGL1987}
Halperin M, Gilbert PR, Lachin JM (1987) Distribution- Free Confidence Intervals for Pr(X1 < X2). Biometrics 43: 71--80.

\item{HMN1982}
Hanley JA, McNeil BJ (1982) The meaning and use of the area under a receiver operating characteristic (ROC) curve. Radiology 143: 29--32.

\item{Hilgers1981}
Hilgers R (1981) On an Unbiased Variance Estimator for the Wilcoxon-Mann-Whitney Statistic Based on Ranks. Biom J 23, 665--661.

{\color{black}
\item{Kruskal1952}
Kruskal, W. H. (1952). A nonparametric test for the several sample problem. The Annals of Mathematical Statistics 23, 525--540.
}

\item{Lehmann1951}
Lehmann EL (1951) Consistency and unbiasedness of certain nonparametric tests. Ann Math Stat 22, 165-179.

\item{Levy1925}
Lévy, P. (1925). Calcul des probabilités. Gauthier-Villars, Paris.

\item{MW1947}
Mann HB,  Whitney DR (1947) On a test of whether one of two random variables is stochastically larger than the other. Ann Math Stat 18: 50--60.

\item{Mee1990}	
Mee RW (1990) Confidence Intervals for Probabilities and Tolerance Regions Based on a Generalization of the Mann-Whitney Statistic. J Am Stat Assoc 85: 793-800

\item{Newcombe2006}
Newcombe, R. G. (2006). Confidence intervals for an effect size measure based on the Mann Whitney statistic. Part 2: asymptotic methods and evaluation. Statistics in Medicine, 25(4), 559--573.

\item{NoPaBr2022}
Nowak CP, Pauly M, Brunner E (2022) The nonparametric Behrens-Fisher problem in small samples.
 https://arxiv.org/pdf/2208.01231

\item{OW1980}
Orban J, Wolfe DA (1980). Distribution-free partially sequential placement procedures. Commun Stat - Theory Methods 9: 883--904.

\item{OW1982}
Orban J, Wolfe DA (1982) A class of distribution-free two-sample tests based on placements. J Am Stat Assoc 77: 666--672

\item{PeMa2019}
Perme MP, Manevski D (2019) Confidence intervals for the Mann-Whitney test. Stat Methods Med Res 28: 3755--3768

\item{Putter1955}
Putter J (1955) The treatment of ties in some nonparametric tests. Ann Math Stat 26: 368-386

{\color{black}
\item{RaWo1979}
Randles, R.H., Wolfe, D.A. (1979). Introduction to the theory of nonparametric statistics. Wiley, New York.
}

\item{Ruy1980}
Ruymgaart FH (1980) A unified approach to the asymptotic distribution theory of certain
midrank statistics. In: Statistique non Parametrique Asymptotique: 1--18. JP Raoult (Ed). Lecture Notes on Mathematics No. 821, Springer, Berlin.

\item{Sen1967}
Sen PK (1967) A note on asymptotically distribution-free confidence intervals for $Pr(X<Y )$ based on two independent samples. Sankhya A 29: 95--102

\item{Shirahata1993}
Shirahata S (1993) Estimate of Variance of Mann-Whitney Statistic. J Jap Soc Comp Stat 6: 1--10

\item{Dantzig1951}
Van Dantzig D (1951) On the consistency and power of the Wilcoxon's two-sample test. Koninklijke Nederlandse Akademie van Wetenschappen, Proc Ser A 54: 1-9

\edes

\end{document}